\documentclass[fleqn,usenatbib]{mnras}

\usepackage[T1]{fontenc}

\usepackage{ae,aecompl}
\DeclareRobustCommand{\VAN}[3]{#2}
\let\VANthebibliography\thebibliography
\def\thebibliography{\DeclareRobustCommand{\VAN}[3]{##3}\VANthebibliography}

\usepackage{graphicx}	
\usepackage{amsmath}	
\usepackage{subcaption}
\usepackage{amssymb}	

\newcommand{\Msunh}{\,{\rm h}$^{-1}$\,\ifmmode M_{\odot}\else $M_{\odot}\,$\fi}
\newcommand{\Mpch}{\,{\rm Mpc}\,\ifmmode h^{-1}\else $h^{-1}$\fi}
\newcommand{\kpch}{\,{\rm kpc}\,\ifmmode h^{-1}\else $h^{-1}$\fi}

\title[Mutual Information between galaxy assembly and dark matter halo assembly]{Exploring the link between galaxy assembly and dark matter halo assembly in IllustrisTNG: Insights from the Mutual Information}

\author[Y. D. Camargo. \& R. A. Casas-Miranda]{
Y. D. Camargo,$^{1}$\thanks{E-mail: ydcamargoc@unal.edu.co}
R. A. Casas-Miranda $^{1}$\thanks{E-mail: racasasm@unal.edu.co}
\\
$^{1}$Departamento de Física, Universidad Nacional de Colombia - Sede Bogotá, Av. Cra 30 No 45-03, CP 111321 Bogotá, Colombia}

\date{Accepted XXX. Received YYY; in original form ZZZ}

\pubyear{2015}

\begin{document}
\label{firstpage}
\pagerange{\pageref{firstpage}--\pageref{lastpage}}
\maketitle


\begin{abstract}
    We employed Mutual Information (MI) analysis to investigate the relationship between galaxy properties and the assembly history of their host dark matter (DM) haloes from the IllustrisTNG simulations. Focusing on central and satellite galaxies with stellar masses between $10^{9} \, - \, 10^{11.5}$ \Msunh, we examined the correlation between halo assembly time and galaxy assembly time, specific star formation rate (sSFR), color $(g-i)$, and galaxy formation efficiency $F_\star$. Our results indicate a strong correlation between $F_\star$ and the halo assembly time for low-mass central galaxies, suggesting a co-evolutionary relationship. In contrast, sSFR and color $(g-i)$ exhibit weaker correlations with halo assembly time, indicating that additional factors should influence these galaxy properties. Satellite galaxies show negligible correlation between their properties and halo assembly time, highlighting the impact of environmental processes on their evolution. We further extended our analysis to cluster observables, including the magnitude gap, the satellite richness, and the distances to the satellites. Although these cluster properties display weak overall correlations with halo assembly time, the richness consistently increases with stellar mass. This trend suggests that richness is more closely linked to formation history in more massive haloes, where satellite accretion dominates the growth of their host DM haloes. These findings establish $F_\star$ as a more sensitive indicator of halo assembly history than colour $(g-i)$, sSFR, or cluster observables, offering new insights into the complex interplay between galaxy evolution and the hierarchical growth of their host dark matter haloes.
\end{abstract}



%


\section{Introduction}
\label{sec:intro} 

The observed galaxies provide us with fundamental information to understand the large-scale structure of the universe. According to the ($\Lambda$CDM) model, these galaxies form and evolve inside Dark Matter (DM) haloes as a result of the non-linear evolution of density perturbations in matter, establishing a close relationship between the galaxies and their host DM haloes. The study of this connection is a very active area of research in cosmology, involving various methodologies and approaches, such as the analysis of galaxy clustering measurements and modelling the formation of structures using numerical simulations. However, the details of this relationship and how they reflect in the observational properties of galaxies are not yet fully understood.

 The relationship between galaxy stellar mass and DM halo mass is complex, as shown by the scatter of the Stellar-to-Halo Mass (SMHM) relation. This scatter suggests that factors beyond stellar mass and halo mass influence the galaxy-halo connection (see \citet{2018ARA&A..56..435W} for a comprehensive review). One of those potential factors might be a relationship between the assembly time of galaxies and the assembly time of their DM haloes. By examining the potential correlation between galaxy assembly times and halo assembly time, we can gain further insight into the complex processes underlying galaxy formation and its interplay with host haloes, as well as how this relationship manifests itself in observational properties such as galaxy clustering. This is particularly relevant from an observational point of view. For instance, \citet{2008ApJ...687..919W} have found that groups with a red central galaxy or a mean colour redder than the galaxy members exhibit stronger clustering than groups of the same mass but with blue central galaxies. This observation implies a correlation between the colour properties of central galaxies and the clustering behaviour of galaxy groups. 
 To further explore this connection, \citet{10.1093/mnras/stt1374} introduced a specific variant of the halo abundance matching (HAM) model, termed the Age Distribution Matching model. In this model, the galaxy colour is monotonically correlated with the halo age at a fixed halo maximum velocity, $V_{\rm max}$. As a result, the oldest haloes (early-formed haloes) contain the reddest galaxies, whereas younger haloes (late assembly haloes) contain the bluest galaxies. The Age Distribution Matching model has been widely adopted in studies of galaxy catalogues to link galaxy colour, as a proxy for halo age, with the properties of simulated dark matter haloes.

 \citet{2014MNRAS.444..729H} found that the Age Distribution Matching model correctly predicts the colour ($g-r$) of central and satellite galaxies, the clustering measure from the projected galaxy two-point correlation function (2PCF) as a function of stellar mass, and the scaling of these colours with the mass of their host halo from the Sloan Sky Digital Survey (SSDS) DR7. In general, the age-distribution matching model relies on the assumption that colour and sSFR serve as proxies for halo age. To study halo assembly bias \citet{2016ApJ...819..119L} assumed that the formation history of haloes can be inferred from the properties of their central galaxies. They considered a monotonic relationship between halo age and the sSFR but found no convincing evidence of assembly bias due to the lack of correlation between the sSFR and the halo formation history. On the other hand, \citet{Lim_2016} proposed using the ratio of the stellar mass of central galaxies to the mass of their DM haloes ($f_c = M_{*,c} / M_h $) as an observational proxy for the halo assembly time. They used the galaxy groups from the SDSS. Their findings indicated that haloes with higher $f_c$ had assembled their masses earlier, and central galaxies are found to be redder and more quenched in star formation. Alternatively, \citet{2017MNRAS.472.2504T} used galaxy group catalogs from the NYU Value-Added Galaxy Catalog to investigate the relationship between halo age and galaxy properties. They defined halo age using the peak halo mass  ($M_{\rm peak}$) as a proxy, which mitigates the impact of splashback haloes. They found that for central galaxies with stellar masses greater than  $10^{10} M_\odot/ h^{-2}$ there is a $\sim 5$ per cent increase in the fraction of quenched galaxies ($f_q$) from low-to-high density environments. This suggests a strong positive correlation between $f_q$ and the environmental density ($\rho$) for models where the halo age aligns with the central galaxies age. However, this correlation weakens for less massive galaxies, indicating a potential inconsistency with the age-matching model in these cases. Their findings imply that halo formation history has a small but statistically significant impact on quenching star formation in high-mass galaxies, whereas the quenching process in low-mass central galaxies is uncorrelated with halo formation history. Recent results from \citet{Berti_2023} suggest that the colour of the galaxy is not correlated with the age of the halo in the LRG regime. They used SHAM and age distribution matching techniques to construct a magnitude-limited galaxy catalogue and subsequently selected a sample of Luminous Red Galaxies (LRGs) from this catalogue by applying the DESI LRG selection criteria. They used a halo sample from the Multidark Planck2 (MDPL2) simulation and determined the ages of the DM haloes in the same way as \citet{10.1093/mnras/stt1374}. %

 Understanding the formation and evolution of large-scale structure requires considering not only the mass of dark matter halos but also their assembly history. This phenomenon, known as \emph{Halo assembly bias}, is predicted in cosmological simulations, e.g. \citep{2005MNRAS.363L..66G,2006ApJ...652...71W,2008MNRAS.389.1419L,10.1093/mnras/sty109}. Furthermore, if galaxy assembly histories are correlated with the assembly histories of their host haloes, one would expect to observe \emph{galaxy assembly bias} in galaxy clustering measurements. This phenomenon has been quantified using N-body simulations, semi-analytic models, and hydrodynamic simulations, e.g. \citet{2007MNRAS.374.1303C,2014ApJ...794...74J,2019MNRAS.484.1133C, 2018ApJ...853...84Z} \citet{2018MNRAS.480.3978A,2020MNRAS.tmp.1844M, 2021MNRAS.508..940M, 2020MNRAS.492.2739X}. However, from an observational point of view, the galaxy assembly bias remains controversial and has produced mixed results mainly because estimating a halo's mass and formation history from observable galaxy properties is a complex process with inherent uncertainties. The difficult problem of detecting halo assembly bias from observations has been tackled with several approaches \citep{2006ApJ...638L..55Y,Lacerna_2014,2016ApJ...819..119L, 2017MNRAS.470.4767B, 2017MNRAS.468.3251D,2017MNRAS.470..551Z, 2018MNRAS.477L...1N}. Despite the difficulties, researchers have employed various techniques to investigate halo and galaxy assembly bias. For instance, \citet{2019MNRAS.488.3143B} proposed a model that links galaxy star formation rates (SFRs) to the assembly history and redshift of their host haloes. They found a stronger correlation between SFR and assembly history in smaller haloes compared to larger ones. Notably, the SFR of large haloes shows more scatter, suggesting a wider range of star formation processes occurring within these massive environments. Their findings align with the concept of halo assembly bias. However, other studies have challenged these findings. Observations using weak lensing have raised concerns about projection effects that mimic halo assembly bias (e.g., \citet{2016PhRvL.116d1301M, 2019MNRAS.490.4945S, 2022arXiv220503277S}). Additionally, the strength of the correlation between galaxy assembly and halo assembly appears to be a debated topic, since some studies using numerical simulations suggest a connection between them, particularly for specific galaxy properties (e.g., \citet{2020MNRAS.tmp.1844M, 2021MNRAS.508..940M}), while other studies that use group catalogues or numerical simulations with HOD models show mixed results (e.g., \citet{2018MNRAS.478.4487T, 2019MNRAS.485.1196Z}). \\

Building upon the limitations identified with current methods that attempt to link observed galaxy properties to simulated halo properties, a deeper understanding of the intricate relationship between these entities is still sought. In this sense, the field of information theory provides us with valuable tools, such as Mutual Information (MI) analysis, originally proposed by \citet{shannon1948mathematical}. The MI captures the relationship between two random variables that are sampled simultaneously. It is widely used in machine learning and deep learning models for classification tasks, feature selection, etc., and has emerged as a powerful method in cosmology that can provide valuable insights. For example, this technique allows for a quantitative exploration of the mutual dependence between galaxy and halo properties, shedding light on the complex interplay of assembly histories and other factors. Previous works such as \citet{2017MNRAS.471L..77P, 2020JCAP...09..039B, 2020MNRAS.497.4077S,2022PhRvD.105j3533L,2023arXiv230704994S} have demonstrated the utility of MI statistics to uncover meaningful correlations in cosmology. Using MI analysis, we can dive deeper into the underlying connections and uncover hidden dependencies that may not be evident with traditional methods. This is because the MI can detect any kind of dependence, unlike Pearson's correlation coefficient $r_{x,y}$, which is a simpler measure that only assesses the linear relationship between two variables $x$ and $y$.\\

Our goal is to shed light on the complex interplay between the formation of galaxies and the assembly history of their host dark matter haloes, which is essential for unravelling the complexities of large-scale cosmic structure. 
The analysis focuses on the potential dependence of galaxy properties on the assembly time of their host dark matter haloes. Specifically, we examine the dependence between galaxy assembly time, specific star formation rate (sSFR), colour $(g-i)$, and galaxy formation efficiency ($F_\star = log_{10}(M_\star / M_h)$) with halo assembly time as a function of stellar mass. Using Mutual Information statistics, we determine the extent to which these galaxy properties can serve as proxies for halo assembly time. Our analysis is based on data from the IllustrisTNG cosmological hydrodynamical simulation, which has been shown to accurately match observations of clustering properties \citep{2018MNRAS.475..676S}.

This paper is organized as follows: In Section \ref{sec:simul}, we provide an overview of the IllustrisTNG simulation and the construction of the galaxy sample. In Section \ref{sec:galaxy_prop} we explore some aspects of the galaxy-halo connection. Using the MI analysis, in Section \ref{sec:times} we will do a comprehensive exploration of the intricate relationships of the galaxy assembly times, the specific Star Formation Rate (sSFR), the colour $(g-i)$, and $F_\star$ with the assembly times of host DM haloes. Finally, in Section \ref{sec:conclu} we discuss our results and in the appendix \ref{sec:ApenMI} we show some details about the MI.

\section{Simulation and Data Samples}
\label{sec:simul}

We used data from the state-of-the-art hydrodynamic galaxy formation simulation The Next Generation (Illustris-TNG)\footnote{\url{https://www.tng-project.org/}}. In particular, we used data of galaxies and DM haloes from the simulation labeled TNG300-1 \citep{2019ComAC...6....2N}, which provides the largest volume with the highest resolution. This simulation was performed with the AREPO code \citep{2010MNRAS.401..791S} which solves for gravitational and magnetohydrodynamic physics and includes subresolution physics to describe star formation, black hole growth, and feedback processes. 
This simulation was performed on a cubic volume of $205$ \Mpch on a side, using  $2500^3$ DM computational particles of mass $3.98 \times 10^7$ \Msunh; the average gas cells have a mass of $7.43 \times 10^6$ \Msunh. The cosmological parameters in the simulation were $\Omega_m=0.38089$, $\Omega_b=0.0486$, $\Omega_\Lambda= 0.6911$, and $h=0.6774$ consistent with Cosmic Microwave Background measurements from the Planck satellite \citep{2016A&A...594A..13P}.

In the IllustrisTNG simulation, the DM haloes are identified using the Friends-of-Friends (FoF) algorithm with a linking length of 0.2 times the mean interparticle separation, while subhaloes are identified using the SUBFIND algorithm \citep{2015MNRAS.449...49R}. The assembly time of the haloes is estimated using the baryonic merger trees built from 100 snapshots ($z= 0 \, \,\text{to} \, \, z= 20$) using the SUBLINK algorithm at the subhalo level \citep{2015MNRAS.449...49R}. The haloes, subhaloes, and merger trees are available in the IllustrisTNG database. To construct the galaxy sample, we use the stellar mass within twice the stellar half-mass radius of each subhalo as the galaxy stellar mass measurement. Our sample only includes galaxies with stellar masses $M_\star > 10^{9}$ \Msunh at $z = 0$, this allows us to have galaxies with a well-resolved formation history; above this threshold the galaxies are resolved with at least $150$ stellar particles. We impose a minimum of $10^3$ galaxies in a mass bin of 0.25 dex in width to have good correlation function measurements. This limitation translates into an upper stellar mass limit of $M_\star=10^{11.5}$\Msunh for our samples. 

To establish the formation times of galaxies and DM haloes, we extract the main branch of the tree by following back every snapshot of the most massive progenitor. The assembly time is defined as the redshift at which the galaxy or DM halo in the branch reaches exactly half of its stellar mass or halo mass at z=0 ($z_{G}$ for the assembly time of the galaxies and $z_{H}$ for the assembly of the DM haloes). For DM haloes, identified from high-resolution N-body simulations, \citet{2011MNRAS.413.1973W} found that the assembly time ($z_H$) is closely related to its internal structure and dynamical properties. This definition of halo formation time has been widely adopted in the literature and is believed to have a significant impact on the properties of galaxies, such as galaxy age, colour, and star formation rate (SFR) \citep{2008MNRAS.389.1419L}. Moreover, in hydrodynamical simulations, \citet{2017MNRAS.465.2381M} and \citet{2020MNRAS.491.5747M} found that scatter in the stellar-to-halo mass (SMHM) relation in central galaxies correlates with the halo assembly time. The DM halo mass is estimated using the total mass enclosed within a sphere whose mean density is 200 times the critical density of the universe at $z=0$. Although the redshift at which a galaxy has assembled half of its current mass ($z_{G}$) is not directly observable, it is possible to estimate this assembly redshift with methods such as carefully analyzing the observed properties of the galaxy's spectral energy distribution. This approach, proposed by \citet{2020MNRAS.497.4262J}, can be used to make inferences about the assembly time of galaxies, providing valuable information about their evolutionary histories.

\section{Galaxy-Halo Mass Connection}
\label{sec:galaxy_prop}
\begin{figure*}
    \centering    
        \includegraphics[width=\linewidth, keepaspectratio]{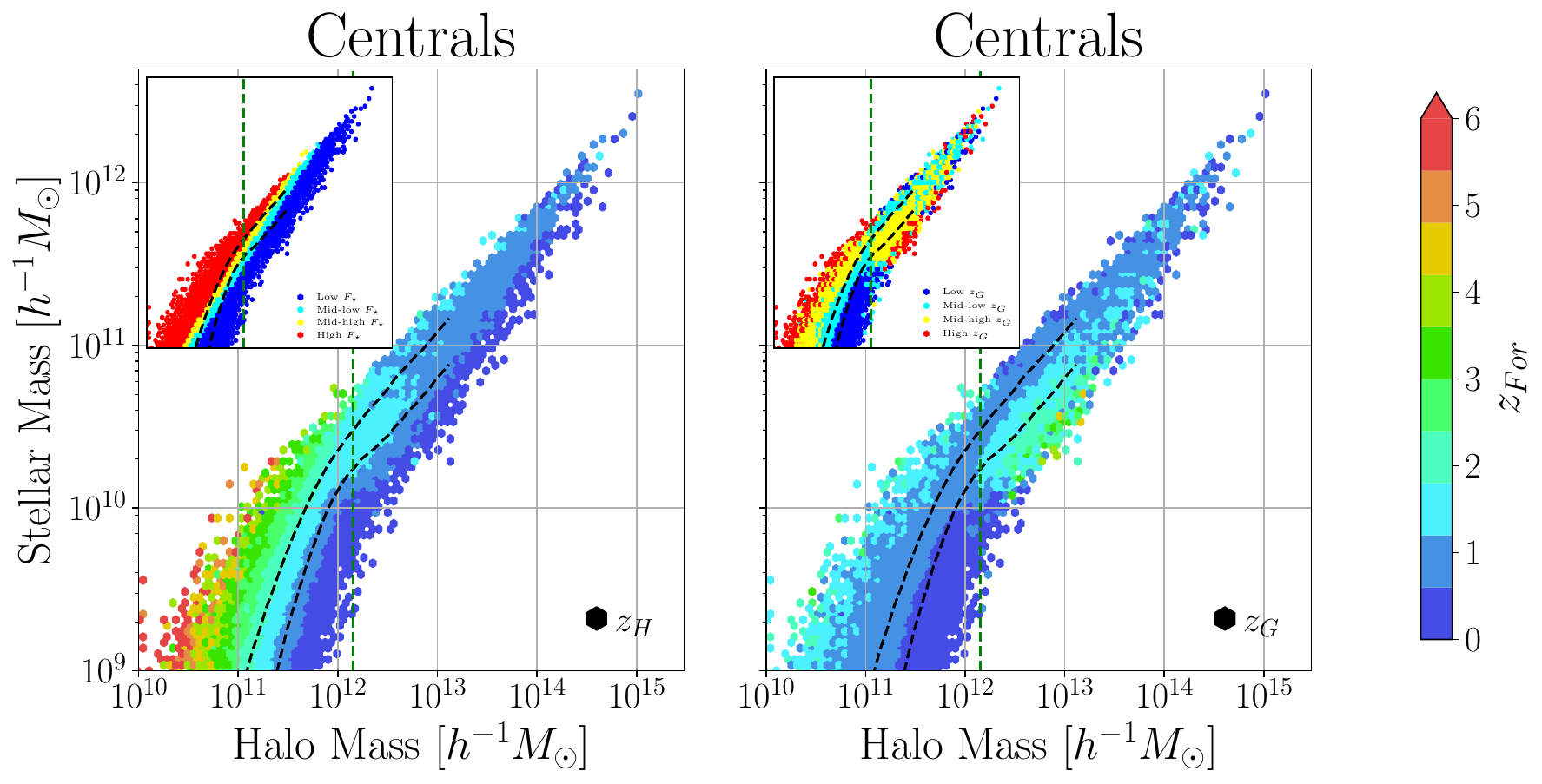}
        
        \includegraphics[width=\linewidth, keepaspectratio]{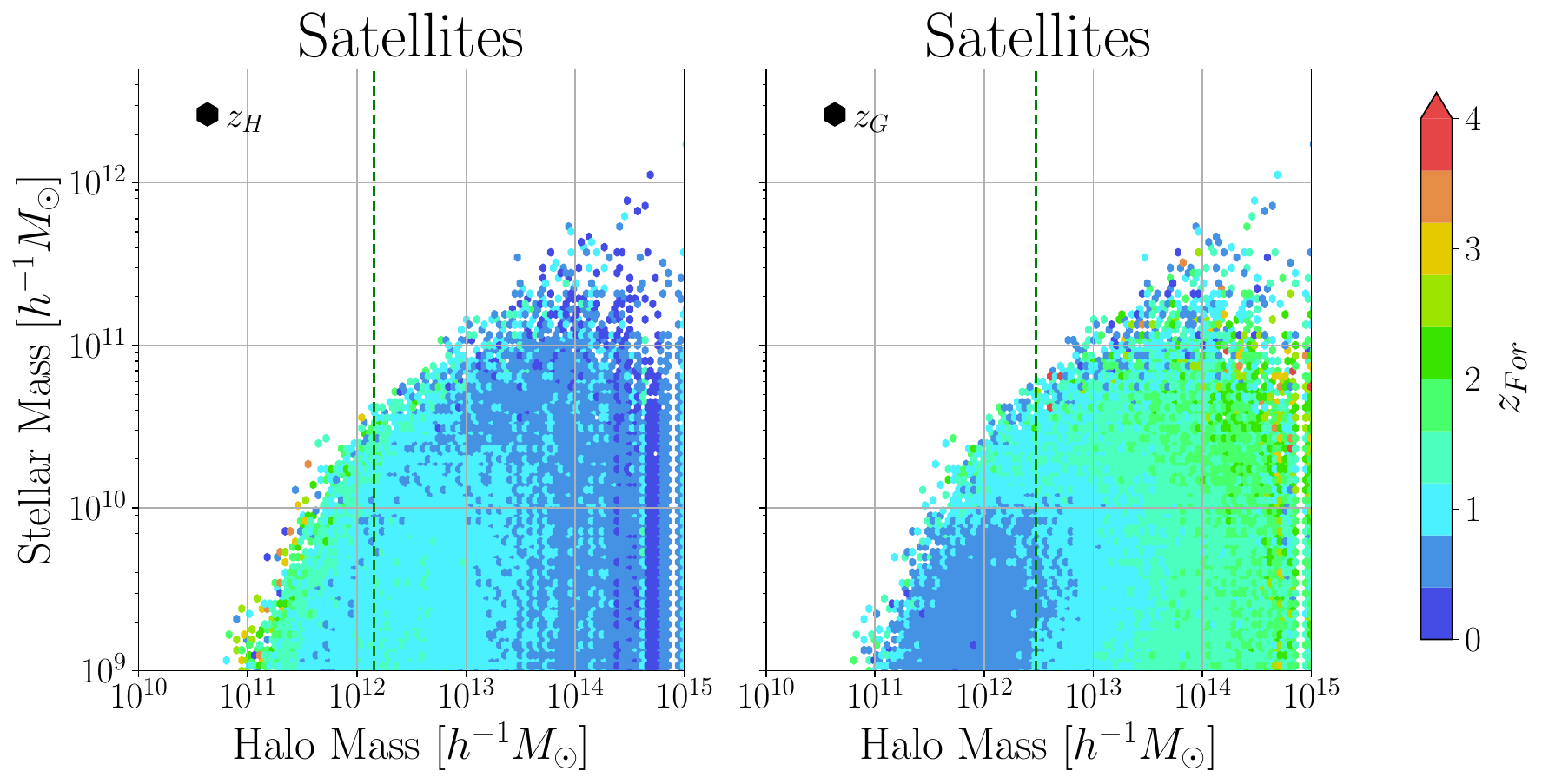}

    \caption{Stellar-to-halo mass relation (SMHM) for central galaxies upper panel and satellite galaxies lower panel. Black dashed curves in the upper panels represent the region within $1\sigma$ of the mean stellar mass at fixed halo mass. Central galaxies show a pronounced relationship between the stellar mass of the galaxies and the mass of their host halo, particularly for haloes with $M_{h} \lesssim 1.44 \times 10^{12}$ \Msunh (show by dashed green lines). The colour scale indicates the assembly times of DM haloes ($z_{H}$) in left panel and assembly times of galaxies ($z_{G}$) in right panel. Right inset figure focuses on the SMHM relation for central galaxies, colour-coded by $z_{G}$ quartiles, left inset figure focuses on the SMHM relation for central galaxies, colour-coded by $F_{\star}$ quartiles. The $x$- and $y$-axes in the inset figures share the same scales as those in the main central galaxy panels.}
    \label{fig:stellar_to_halo}
\end{figure*}

A cornerstone of galaxy formation theory, the stellar-to-halo mass (SMHM) relation quantifies the efficiency of star formation and reflects the influence of stellar feedback and feedback from Active Galactic Nuclei (AGN). This relationship has been extensively studied in hydrodynamic simulations and semi-analytical models (SAM) (e.g. \citet{2014MNRAS.444.1518V,2016ApJ...833....2G, 2017MNRAS.465.2381M, 2018MNRAS.480.3978A, 2018ApJ...853...84Z,  2019MNRAS.490.5693B, 2019MNRAS.488.3143B, 2020MNRAS.498.1839X,2021NatAs...5.1069C}).

To examine the trends in the assembly times of galaxies and their host DM haloes, Figure \ref{fig:stellar_to_halo} shows the SMHM relation for central and satellite galaxies from the TNG300-1 simulation. The colour coding represents the range of assembly times of galaxies (right panels) and the range of assembly times of their DM haloes (left panels). For central galaxies (upper panels), a clear trend emerges: more massive galaxies are hosted in more massive DM haloes, demonstrating a relatively tight relationship between stellar and halo mass. This relationship exhibits a steep slope around $M_h \sim 1.44 \times 10^{12}$ \Msunh (dashed green line). At higher halo masses, this slope becomes less pronounced as AGN feedback and mergers become more important (see Appendix \ref{sec:ApenAGN}), which reduces the efficiency of star formation and ultimately leads to quenching it in central galaxies hosted by these haloes (e.g., \cite{2017MNRAS.465.2381M, 2018MNRAS.480.3978A}, \cite{2018ARA&A..56..435W}). This trend is illustrated in the left inset panel of Figure \ref{fig:stellar_to_halo}, where the SMHM relation is colour-coded by $F_\star$ quartiles, offering a deeper understanding of the efficiency of star formation across the halo mass ranges. Haloes with masses $M_h \lesssim 1.44 \times 10^{12}$ \Msunh exhibit the highest values of $F_\star$ (red regions), reflecting peak efficiency in converting gas into stars. This indicates that these haloes have entered a regime in which their deeper gravitational potential wells retain and accrete gas more efficiently, overcoming feedback effects.

In the upper left panel (colour-coded by $z_H$), the SMHM relation shows a negative correlation between the $z_H$ and the halo mass, indicating that less massive haloes tend to assemble earlier, in alignment with the hierarchical formation scenario. Furthermore, at a fixed halo mass, more massive central galaxies tend to reside in DM haloes with earlier assembly times. This trend is consistent with the findings of \citet{2018MNRAS.480.3978A} in hydrodynamical cosmological simulations and \citet{2018ApJ...853...84Z} in semi-analytical models (SAM), who found that at fixed halo mass (below $10^{12}$\Msunh), central galaxies in early-formed haloes within denser environments tend to be more massive and strongly clustered, reflecting halo assembly bias.

The results for $F_\star$ reveal that at fixed $M_\star$, $F_\star$ declines with increasing $M_h$,  showing that the less massive haloes are more efficient at forming stars. For halos in the mass range $10^{11.5} \lesssim M_h \lesssim 10^{12}$, which host central galaxies with stellar masses below $10^{10}$ \Msunh, $F_\star$ decreases significantly (blue regions) reflecting the dominance of supernova feedback and reionization effects, which inhibit gas retention and star formation. In contrast, for DM haloes with $M\gtrsim 1.44 \times 10^{12}$ \Msunh, $F_\star$ remains low due to the impact of AGN feedback, which reduces the efficiency of star formation in these massive systems.

A focus on the $z_G$ (upper right panel of Figure \ref{fig:stellar_to_halo}) reveals that central galaxies in low-mass haloes ($M_h \lesssim 1.44 \times 10^{12}$ \Msunh, dashed green line) tend to have earlier assembly times as their stellar masses increase. It is noteworthy that central galaxies with stellar mass below $\sim 10^{10.20}$ \Msunh and hosted in DM haloes below $\sim 10^{11.5}$ \Msunh both exhibit early assembly times. Since these haloes are more strongly clustered, the net effect is a stronger clustering of the galaxies -a phenomenon known as galaxy assembly bias-. Conversely, more massive central galaxies ($M_\star \gtrsim 10^{10.20}$ \Msunh) tend to be hosted in DM haloes that assembled more recently compared to their stellar assembly times. This indicates that halos with masses above $\sim 1.44 \times 10^{12}$ \Msunh continue to grow through hierarchical mergers, while star formation in their central galaxies ceases earlier. To validate this trend, the right inset in Figure \ref{fig:stellar_to_halo} shows the SMHM relationship by quartiles of $z_G$. This observation is consistent with the \emph{downsizing effect} (see \citet{2008MNRAS.389.1419L} for a comprehensive discussion), where massive galaxies formed stars more efficiently at higher redshift. In such cases, the assembly time of the host halo does not necessarily align with the stellar assembly time of the galaxy. Instead, it reflects a distinct formation process: Although the halo assembled more recently, the galaxy itself probably formed its stars earlier and more efficiently, prior to AGN feedback becoming a dominant factor and mergers becoming more significant (see Appendix \ref{sec:ApenAGN}) \citep{10.1093/mnras/stw2735, Naree_2023}. The trends in $F_\star$ complement the trends observed in $z_H$ and $z_G$. For instance, more massive central galaxies in halos with $M_h \lesssim 10^{12}$ \Msunh often exhibit high $F_\star$, corresponding to efficient star formation and relatively early assembly times. Conversely, massive halos ($M_h \gtrsim 1.44 \times 10^{12}$ \Msunh) show lower $F_\star$, while their central galaxies tend to exhibit earlier stellar assembly compared to their halo assembly times,  consistent with the downsizing phenomenon. 

\begin{figure*}
    \centering
    \includegraphics[width=1.8\columnwidth]{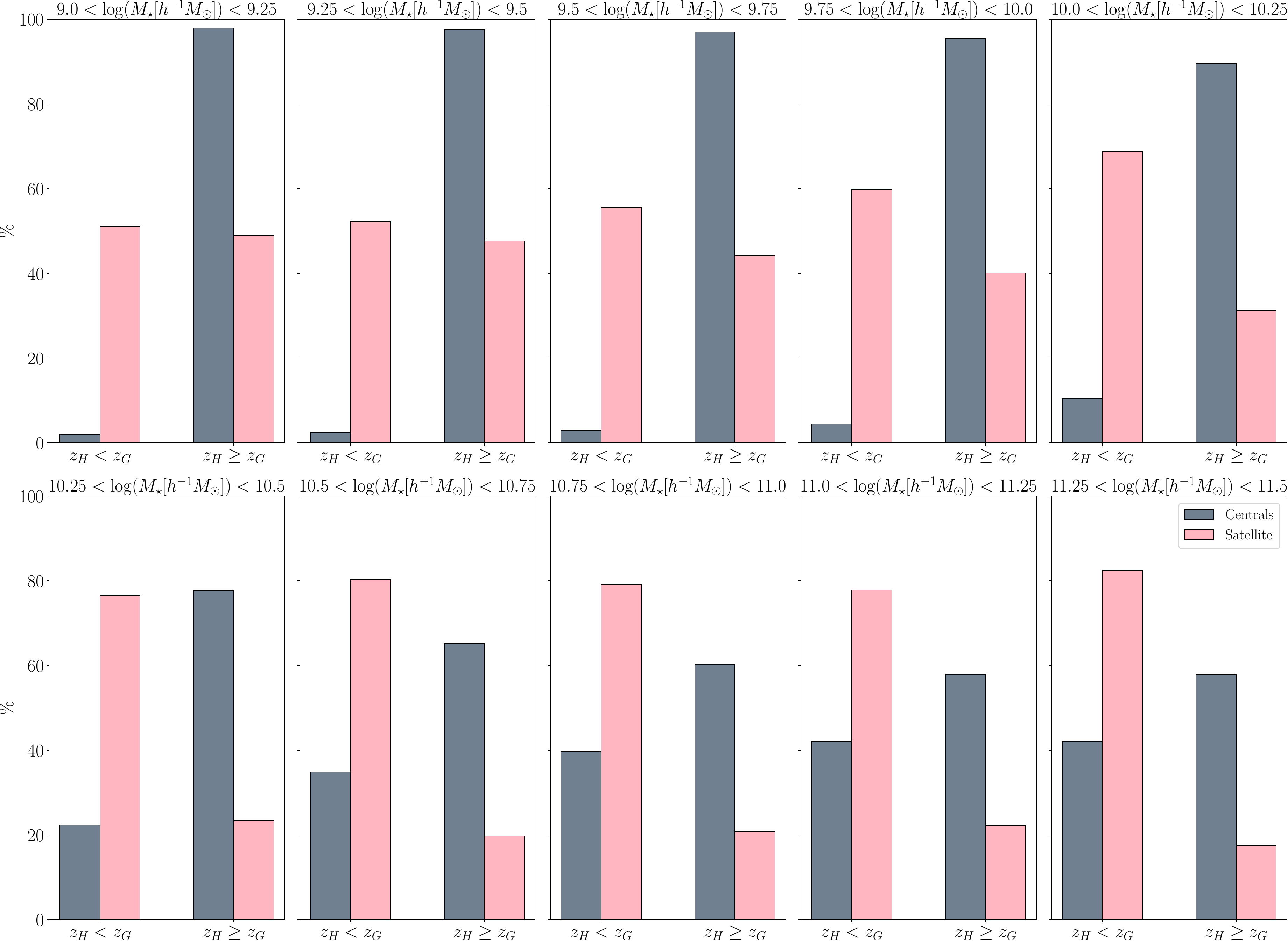}
    \caption{Comparative histograms of $z_{\text{G}}$ and $z_{\text{H}}$ reveal intriguing trends. For central galaxies with stellar masses exceeding $10^{10.25}$ \Msunh a small percentage (2-11\%) exhibit early galaxy formation compared to their host halo formation. The fraction of galaxies with earlier formation times than their host halos ($z_H < z_G$) increases with galaxy stellar mass for both central and satellite galaxies.}
   \label{fig:Histo_Zfor}
\end{figure*}

Figure \ref{fig:Histo_Zfor} provides a more detailed view of this phenomenon, comparing the $z_G$ of central and satellite galaxies with the $z_H$ of their host DM haloes in two cases: $z_H < z_G$ and $z_H \geq z_G$. Notably, for more massive central galaxies, $M_\star > 10^{10.25}$ \Msunh, a significant percentage (22$\%$ to 42$\%$) are hosted in DM haloes that assembled more recently than the galaxies themselves. These results show that galaxies that exhibit a significant downsizing effect are typically hosted in DM haloes with $M_h \gtrsim 1.44 \times 10^{12}$ \Msunh. This observation aligns with findings by \citet{2020MNRAS.499.4748M}, who found that in massive haloes, galaxies form efficiently at high redshift, before the halo mass becomes too large, suppressing late-time star formation and resulting in lower specific star formation rates (sSFR). For central galaxies below this limit ( $M_h \sim 1.44 \times 10^{12}$ \Msunh), the percentage of galaxies showing $z_H < z_G$ ranges from 2$\%$ to 11$\%$. Consequently, the galaxies exhibiting $z_H < z_G$, indicative of late halo assembly compared to the galaxy’s stellar assembly, also tend to show lower sSFRs and redder colours, further supporting the downsizing scenario.
For satellite galaxies, Figure \ref{fig:stellar_to_halo} shows that the SMHM about DM haloes spanning almost three orders of magnitude in halo mass, highlighting the differences in evolution paths of satellite and central galaxies are subject. The evolution of central galaxies is closely tied to the mass of their host haloes, whereas satellite galaxies are more influenced by the properties of their subhaloes. This population of galaxies and their subhaloes is influenced by a specific environmental process during their infall into larger host haloes, such as ram pressure stripping, tidal interactions, and strangulation \citep{2022MNRAS.tmp..806N}. Figure \ref{fig:stellar_to_halo} also reveals that satellite galaxies with stellar masses up to $\sim 10^{10}$ \Msunh residing in DM haloes with masses up to $\sim 3\times 10^{12}$ \Msunh tend to exhibit extended star formation histories, consistent with a late assembly history for these systems. This aligns with the findings of \citet{2020MNRAS.499.4748M}, who showed that low-mass haloes typically accrete less massive satellites, resulting in a low ex situ fraction. In contrast, in massive haloes, accreted satellites often have stellar masses comparable to the central galaxy, leading to higher ex situ fractions. The extended star formation histories observed in low-mass satellites in low-mass haloes are likely tied to their low ex situ fractions, where in situ star formation dominates over mergers.

In terms of assembly times, the satellite galaxies also show a different history with respect to their host DM haloes. The SMHM relation and Figure \ref{fig:Histo_Zfor} show that the phenomenon where $z_H < z_G$ is even more pronounced in satellite galaxies. For lower stellar masses ($M_\star \lesssim 10^{10.25}$ \Msunh), the fraction of satellite galaxies with $z_H< z_G$ reaches 69$\%$, while for more massive satellite galaxies, this fraction increases significantly, reaching up to 83$\%$. This stronger effect highlights the distinct evolution tracks between centrals and satellites. Satellite galaxies often form in smaller halos early in the universe, which are later accreted into larger haloes. Although the stellar components of these galaxies may have formed earlier, their host halos continued to assemble mass over longer periods, resulting in the observed trend $z_H< z_G$. After accretion, satellite galaxies undergo environmental quenching processes such as ram-pressure stripping, harassment, or strangulation, which suppress star formation and lead to a predominantly passive satellite population \citep{2022MNRAS.tmp..806N}. This evolutionary path explains the clearer separation between the formation redshift of the stellar component and the DM halo in the satellites, contributing to the pronounced $z_H < z_G$ trend. For both central and satellite galaxies, the percentage of $z_H < z_G$ increases with stellar mass.\\

Building upon these findings, a natural question emerges: Is there a correlation between galaxy assembly times or other observational proxies with the assembly times of the haloes that host them? Considering this possibility, in our subsequent analysis we investigate the relation between the assembly time of galaxies, the colour $(g-i)$, the sSFR, and the galaxy formation efficiency $F_\star$, with the assembly time of their host DM haloes using the Mutual Information statistics to unravel a potential connection between galaxies and DM haloes.

\section{Mutual Information Analysis of Galaxy Assembly times, $\MakeLowercase{s}SFR$, colour ($\MakeLowercase{g-i}$) and $F_\star$, with their Host DM haloes}
\label{sec:times} 

\begin{figure*}
    \centering
    \includegraphics[width=1.8\columnwidth]{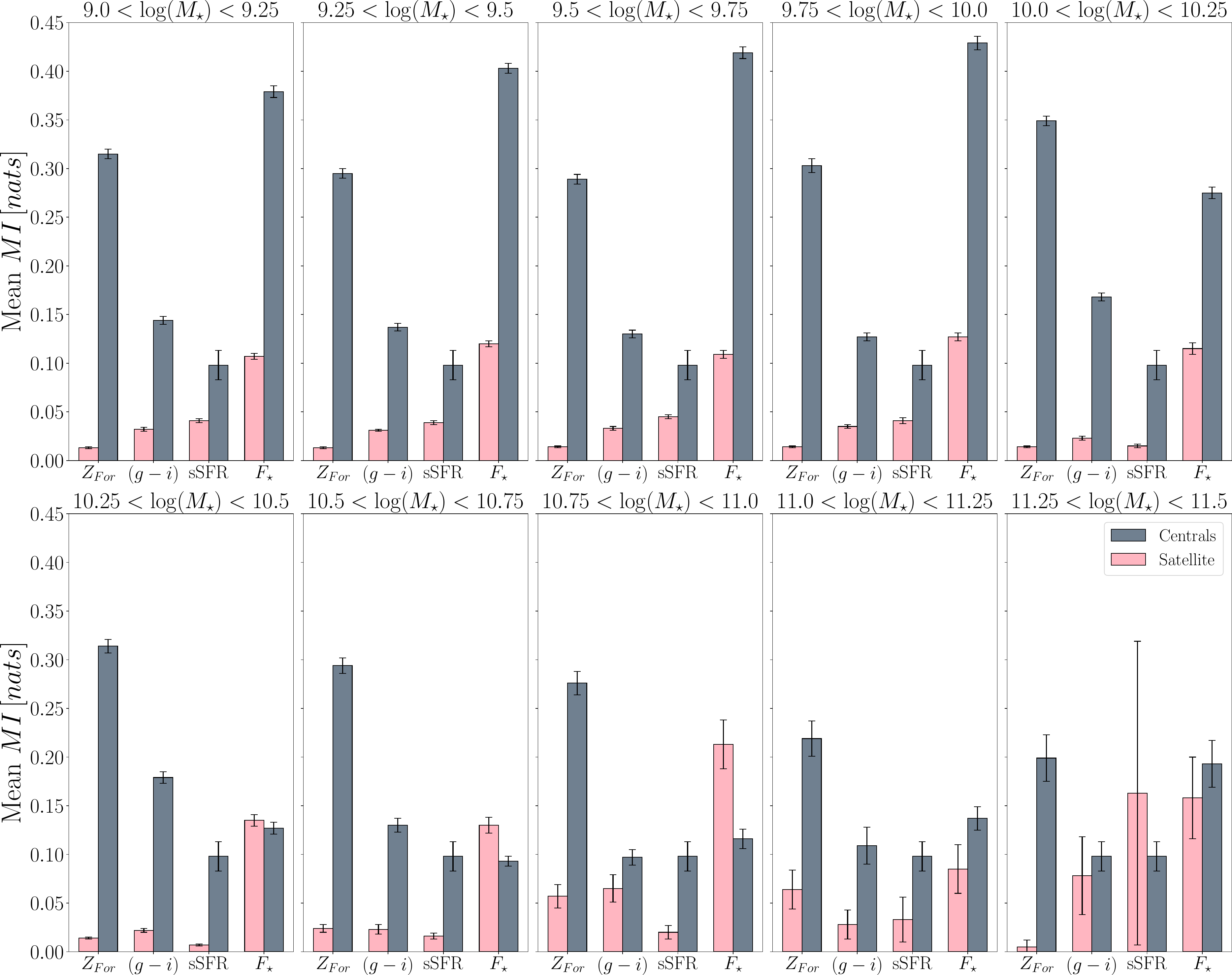}
    \caption{MI quantifying the relationship between galaxy properties (formation time, sSFR, color $(g-i)$, and $F_\star$) and host dark matter halo formation time, separated by central and satellite galaxies, and binned by stellar mass.
    \label{fig:MI_Bar}}
\end{figure*}

Traditionally, correlation analysis methods like Pearson's correlation coefficient have been used to measure the relationship between galaxy properties and the assembly time of their host dark matter haloes. However, these methods are limited in scope, as they only detect linear relationships and may fail to capture the intricate and non-linear nature of these interactions. In this section, we employ the Mutual Information (MI) statistics to obtain a more comprehensive understanding of how galaxy properties are influenced by the assembly history of their host haloes. Specifically, we analyse the galaxy assembly time ($z_G$), the specific star formation rate (sSFR), the colour (g-i), and the galaxy formation efficiency ($F_\star$) to explore potential dependencies on halo assembly time. This approach enables a more complete perspective by quantifying the information shared between variables, providing new insights into the complexities of galaxy formation processes.

For our analysis, we use the \emph{GMM-MI} package\footnote{\url{https://github.com/dpiras/GMM-MI}} \citep{Piras23}. This Python package combines Gaussian Mixture Models (GMMs) with the bootstrap technique (see Appendix \ref{sec:ApenMI} for some details). GMMs is a probabilistic model used to describe data sets generated from a combination of multiple Gaussian distributions. This capability enables GMM-MI to model the underlying distributions of the data, leading to a more robust estimation of the join probability function. The bootstrap technique is then applied to estimate the distribution of the MI and compute associated uncertainty measures.  Combining these two approaches, GMM-MI provides a powerful tool to model complex data structures and capture the shared between variables. GMMs are first applied to individual galaxy and halo properties to identify one-dimensional Gaussian components, and then they are fitted to the joint distribution of $x$ and $y$ to identify two-dimensional Gaussian components. 

\begin{figure*}
    \centering
    \includegraphics[width=0.9\columnwidth]{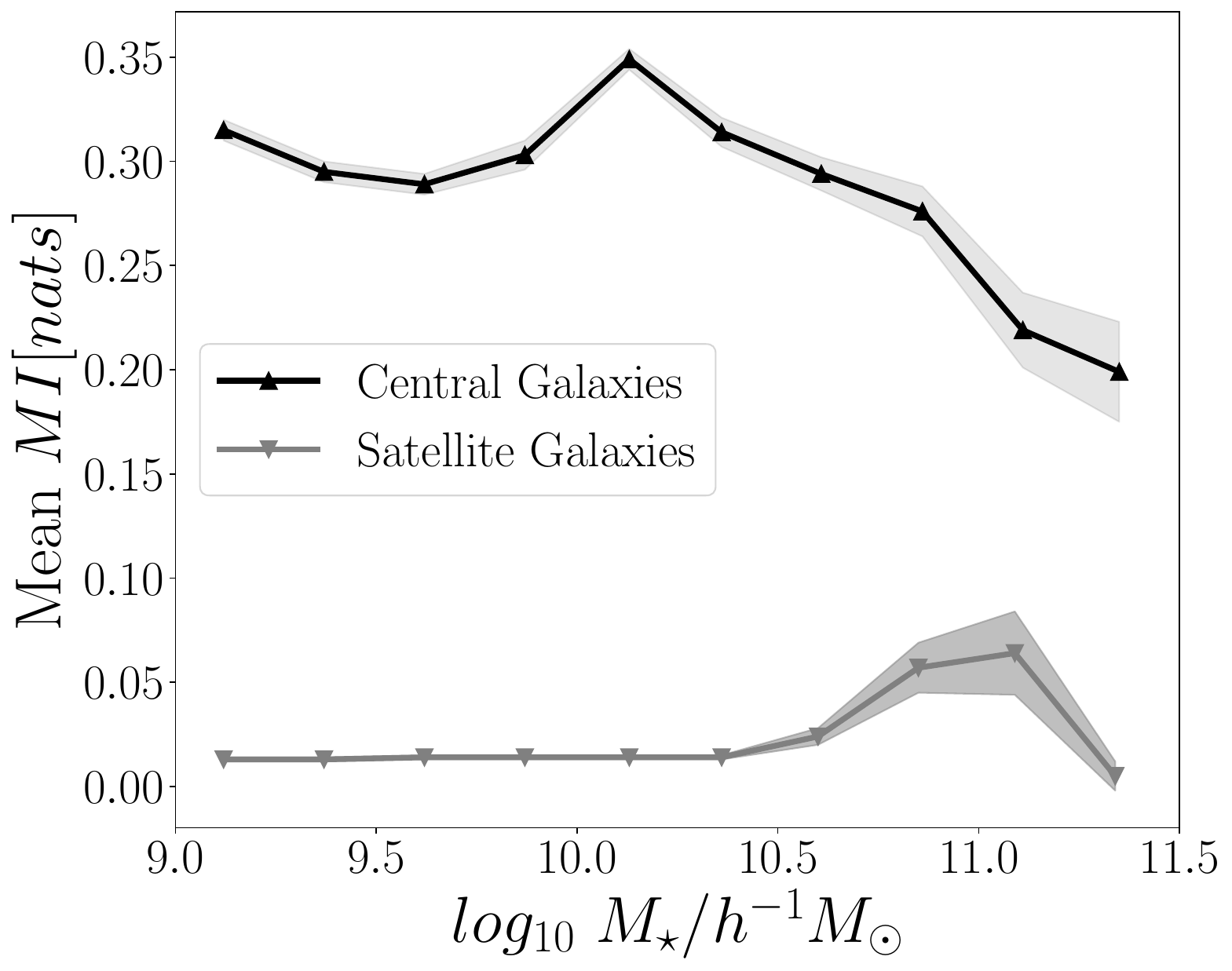}
    \includegraphics[width=0.9\columnwidth]{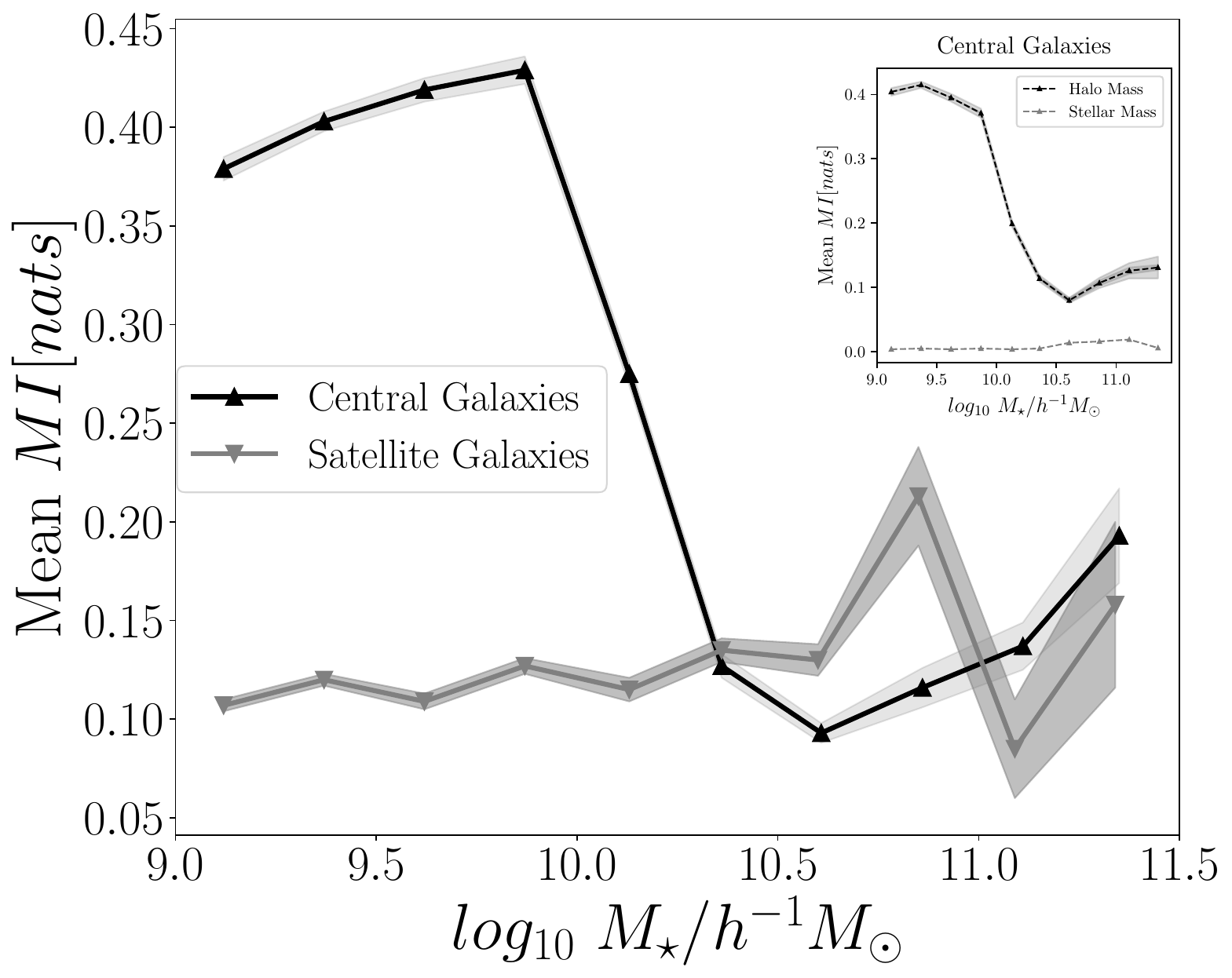}
    \caption{MI as a function of stellar mass of central/satellite galaxies and their host DM haloes. 
Left panel: MI between the formation time of galaxies and the formation time of their DM haloes. Shows that the maximum value is $\sim 0.35\, nats$ for the central galaxies with stellar masses up to $M_\star \lesssim 10^{10.25}$ \Msunh. Right panel: MI between $F_\star$ and the $z_H$. The MI between $F_\star$ and the $z$ assembly of the host halos shows a strongest correlation than the assembly time of galaxies and their host halos. The MI is close to zero for satellite galaxies. The inset shows the MI between galaxy/halo mass and $z_H$ for central galaxies as a function of stellar mass.}
    
    \label{fig:Entropy}
\end{figure*}

Figure \ref{fig:MI_Bar} quantifies the relationships between the variables under study and reveals several important trends. For central galaxies, the MI between colour $(g-i)$ and halo assembly time suggests a weak but discernible correlation ($MI=0.14-0.18 \, nats$) for galaxies with $M_\star \lesssim 10^{10.5}$ \Msunh. In contrast, the MI between sSFR and halo assembly time is even weaker, indicating that sSFR is not a significant predictor of halo assembly time. These findings suggest that neither colour $(g-i)$ nor sSFR are robust proxies for inferring the assembly times of DM haloes, especially for central galaxies. For satellite galaxies, the MI values for both colour $(g-i)$ and sSFR remain close to zero across all mass ranges, indicating negligible correlation with halo assembly times. However, the correlation between $F_\star$ and $z_H$ stands out, particularly for galaxies with $M_\star \lesssim 10^{10.2}$ \Msunh. This likely reflects the recent accretion of low-mass galaxies into their host haloes, leading to extended star formation histories. Although this result suggests a potential connection, further investigation is required to solidify this interpretation. For more massive satellite galaxies, large error bars hinder a confident assessment, which warrants additional analysis to confirm these trends.


Figure \ref{fig:Entropy} provides a more detailed view of the MI between the assembly time of galaxies with $z_H$ (left panel) and $F_\star$ with $z_H$ for the central and satellite (right panel). For central galaxies, the MI with assembly time increases from $0.32\, nats$ to a peak of $0.35 \, nats$ at a characteristic stellar mass of $M_\star \sim 10^{10.25}$ \Msunh (characteristic stellar mass). These findings indicate a tighter correlation between the growth of central galaxies and their host haloes at lower stellar masses, probably because in this mass regime, the halo assembly history strongly dictates the availability of gas and the timing of star formation in the central galaxy. This correlation weakens at higher masses, dropping to $0.2 \,nats$. Satellite galaxies, however, exhibit negligible MI across all stellar mass bins. The right panel of Figure \ref{fig:Entropy} reveals a stronger correlation between $F_\star$ and $z_H$ compared to the assembly times alone. For central galaxies with stellar mass up to $M_\star \sim 10^{10}$ \Msunh, the MI increases from $0.38\,nats$ to $0.43 \, nats$. This trend continues for even more massive galaxies up to $M_\star \sim 10^{10.25} $ \Msunh. This may be attributed to the fact that $F_\star$ directly captures the integrated effects of star formation and feedback processes, which are closely linked to the growth of the halo itself. The tighter correlation of $F_\star$ with $z_H$ suggests that the efficiency of star formation in the central galaxy retains a clearer imprint of the halo assembly time than other properties of the galaxy. This aligns with the idea that, in lower-mass halos, central galaxy growth is more closely tied to the halo's gas accretion and feedback regulation, which are directly related to $z_H$. Beyond this stellar mass range, large error bars make a confident assessment difficult. 

To further investigate the drivers of the observed correlation between $F_\star$ and $z_H$ (see the inset of Figure \ref{fig:Entropy}), we explored the relationship between stellar mass, halo mass, and halo assembly time. Although the MI between stellar mass and halo assembly time was negligible for all stellar mass bins, a significant correlation was found between halo mass and $z_H$ for central galaxies. Specifically, our analysis reveals that for stellar masses in the range of $10^{9} - 10^{10.10}$ \Msunh, the MI is significant, while for halos hosting galaxies of $10^{10} - 10^{10.25}$ \Msunh, a discernible correlation persists. These  results indicate that these galaxies tend to reside in halos that assembled earlier compared to those that host more massive galaxies. The observed correlation between $F_\star$ and $z_H$ for lower-mass galaxies appears to be primarily driven by the correlation between the halo mass and the halo assembly time in this mass range. For more massive central galaxies, the decline in MI between $F_\star$ and $z_H$ is associated with the increasing influence of AGN feedback and merging processes. As shown in the figures \ref{fig:AGNFeedback} and \ref{fig:Mergers}, these processes play crucial roles in shaping the evolution of massive galaxies. Figure \ref{fig:AGNFeedback} illustrates the evolution of black hole accretion activity, quantified as $log_{10}(\dot{M})/log_{10}(M_\star)$, where $\dot{M}$ is the sum of the instantaneous accretion rates of all black holes in a subhalo. This ratio is plotted alongside the stellar mass growth of galaxies as a function of $z$. Figure \ref{fig:Mergers} represents the number of merger events as a function of $z$ for central galaxies. These results collectively highlight the role of AGN feedback and mergers in regulating the stellar mass growth and star formation activity in massive galaxies. Although less massive galaxies are more closely linked to halo assembly time, the evolution of more massive galaxies is dominated by these processes.\\

\begin{figure*}
    \centering
    \includegraphics[width=1.8\columnwidth]{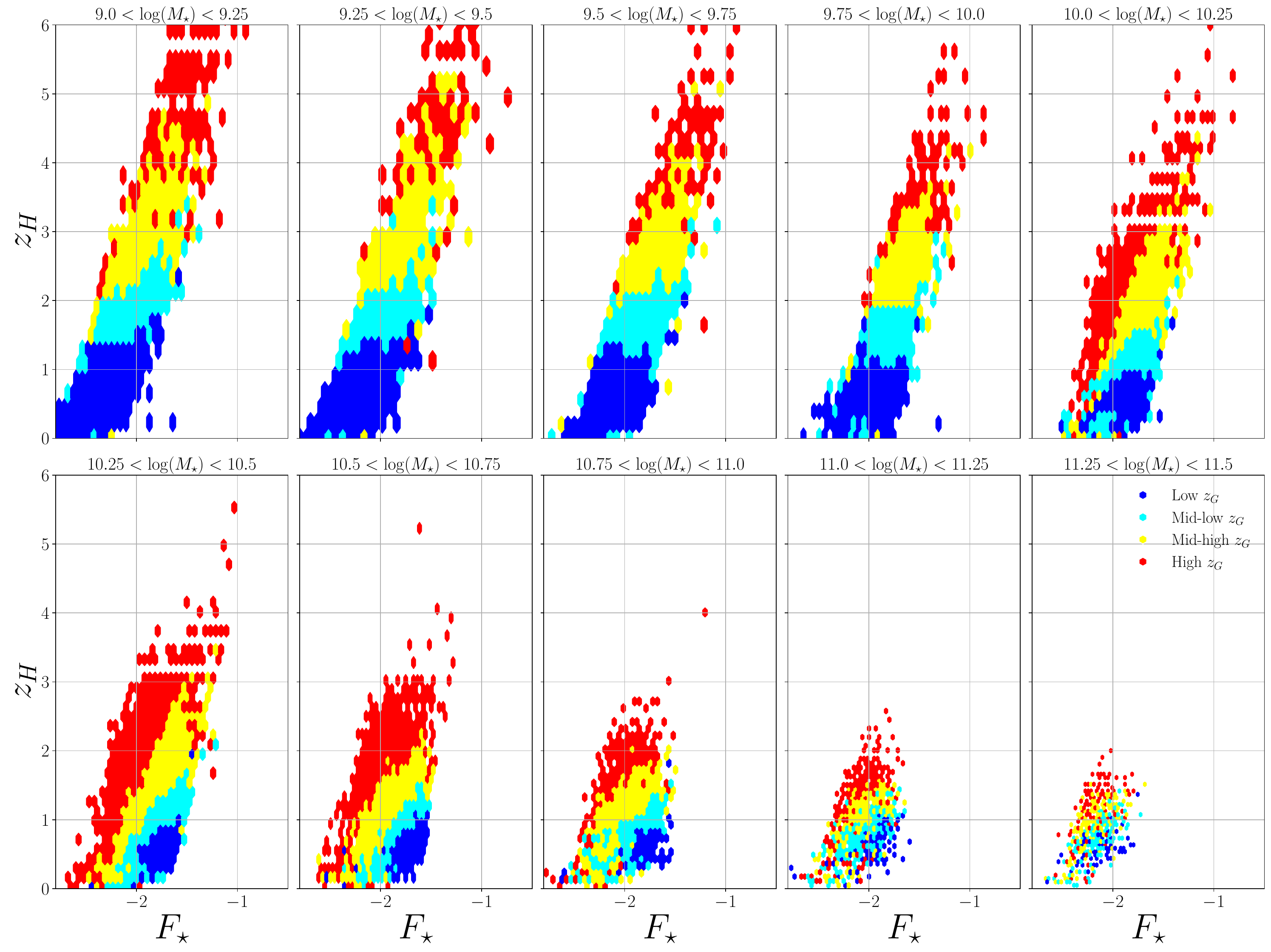}\\
    \caption{Hexbin plot of central galaxies for $z_H$ as a function of $F_\star$, colour-coding by $z_G$ quartiles. A strong/moderate linear correlation between $F_\star$ and $z_H$ to central galaxies with stellar mass up to $10^{10.25}$ \Msunh}
    \label{fig:ZFor}
\end{figure*}

 Motivated by the trends observed in Figure \ref{fig:Entropy} for central galaxies, especially the stronger correlation between $F_\star$ and $z_H$, Figure \ref{fig:ZFor} shows a positive linear correlation (with Pearson coefficient $0.8 \leq r_{x,y} \leq 0.6$) for central galaxies up to the characteristic stellar mass of $M_\star \sim 10^{10.25}$ \Msunh. This indicates that star formation in these halos is closely related to the time of formation. In contrast, the correlation between $F_\star$ and $z_H$ declines rapidly for the most massive galaxies, suggesting that $F_\star$ is less influenced by $z_H$ and more by internal halo processes. Regarding the correlation between these variables and $z_G$, the colour coding reveals that low-mass DM haloes with early assembly times that host low-mass galaxies form their stars early, leading to higher $F_\star$.  This occurs because these halos have more time to form stars \citep{2018ApJ...853...84Z}, before the feedback process limits star formation.
 
 In accordance with Figure \ref{fig:stellar_to_halo}, Figure \ref{fig:ZFor}, binned by stellar mass, highlights how $F_\star$ is connected to both halo formation time and galaxy formation time, showing how these relations evolve with stellar mass. For instance, as stellar mass increases, galaxies with high $z_G$ appear to become more prominent in the context of the diagram. For low stellar mass, the peak in $F_\star$ is observed at $M_\star \lesssim 10^{10.25}$ ($M_h \sim 10^{12}$ \Msunh). For more massive central galaxies, the correlation between $F_\star$ and $z_H$ decreases rapidly, reflects a less structured relationship. Although the large error bars in this mass range make a definitive conclusion challenging, we observed that for massive central galaxies ($10^{11.0}\lesssim M_\star$(\Msunh) $\lesssim 10^{11.5}$), the correlation between $F_\star$ and $z_H$ is moderate, likely due to mergers triggering bursts of star formation. However, these bursts rely on recycled gas, which is less efficient at forming stars compared to pristine gas. This reduced efficiency arises because the recycled gas is typically heated by feedback processes or during mergers, limiting its ability to cool and form stars. The observed weakening correlation at higher stellar masses might also be linked to the role of the CircumGalactic Medium (CGM) acting as a gas reservoir, but further studies are required to confirm this connection.\\


To contextualize these findings, we compare our results with those of \citet{Lim_2016}, who proposed the ratio between the stellar mass of the central galaxy and the mass of its host halo ($f_c$ ) as a proxy for the assembly time of the halo. While \citet{Lim_2016} demonstrated that $f_c$ correlates with galaxy properties such as colour ($g-r$), star formation rate, and size, our study builds upon this approach using MI to explore a wider range of galaxy properties. By quantifying the strength of these relationships, MI provides a more detailed and nuanced perspective on the interplay between halo assembly time and galaxy properties across different stellar mass bins. Our findings align with \citet{Lim_2016} in terms of the general trends observed for these properties with $f_c$, including the downsizing effect observed in central galaxies. However, our results extend these insights by highlighting a stronger correlation between $F_\star$ and the halo assembly time compared to colour $(g-i)$ or sSFR. This trend is particularly evident for central galaxies with stellar masses up to $M_\star \lesssim 10^{10.25}$ \Msunh, suggesting that $F_\star$ might be a more sensitive indicator of halo formation history.

Our findings with respect to the difficulty in establishing a robust correlation between galaxy properties and halo assembly time align with studies that have explored the connection between galaxy properties and halo formation history such as \cite{2016ApJ...819..119L}, who did not find strong evidence for assembly bias using sSFR and star formation history as proxies for halo formation time. Instead, studies such as \citet{2019MNRAS.490.2139R, 2019MNRAS.487.5764T} have highlighted the role of AGN feedback and galaxy merges in shaping the properties of the galaxy.  This suggests that identifying robust observational indicators of halo assembly history remains a challenging task; other factors, such as AGN feedback, are also crucial in shaping the observed trends.

 These findings underscore the complex relationships between galaxy properties and halo assembly time. While colour $(g-i)$ and sSFR demonstrated have limited utility as proxies for halo assembly time, the strong correlation between $F_\star$ and halo assembly times for central galaxies highlights its potential as a more reliable indicator of halo formation history.


\subsection{Mutual Information Analysis for Cluster Observables}
\begin{figure*}
    \centering
    \includegraphics[width=1.8\columnwidth]{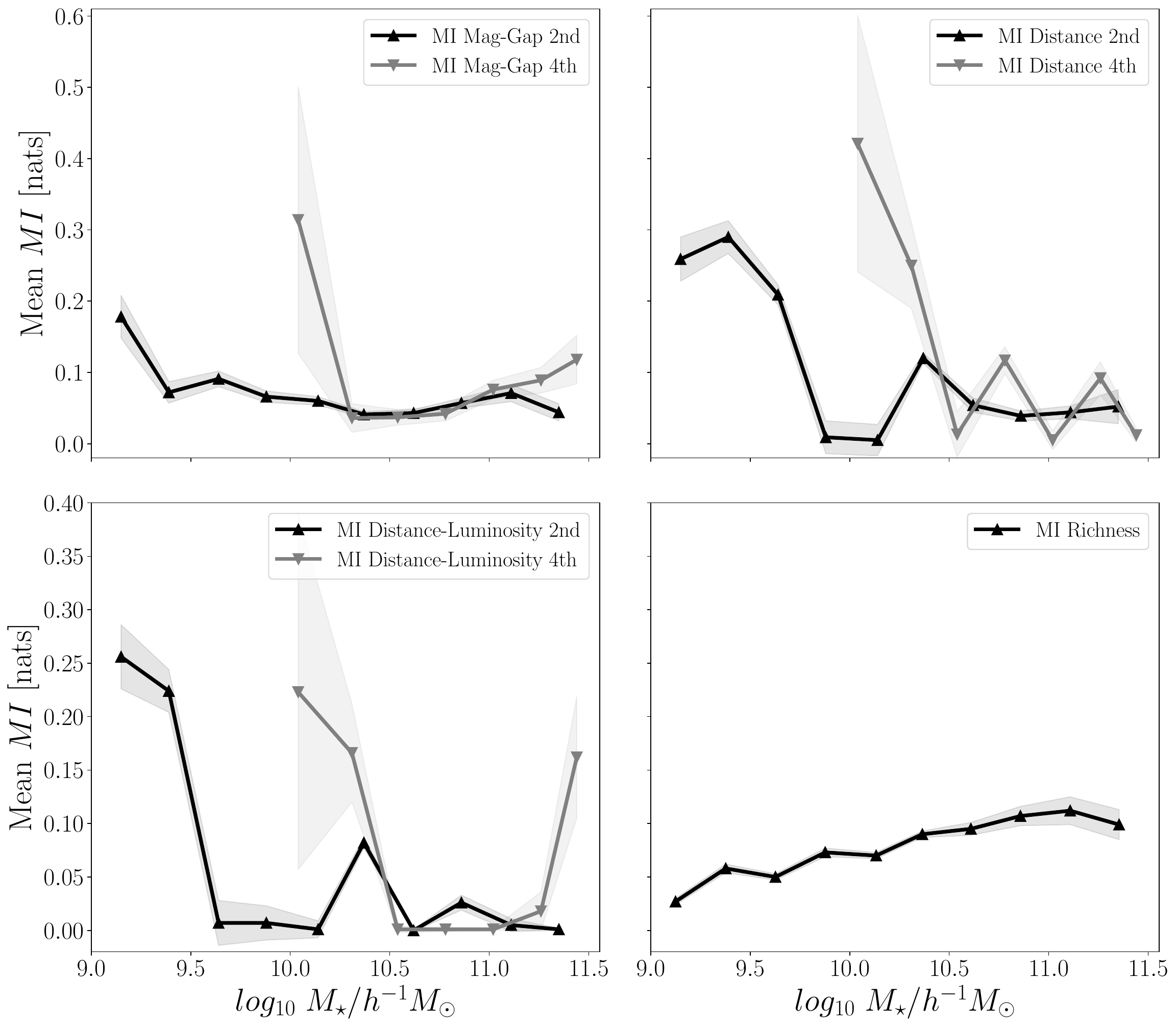}
    \caption{Mutual Information (MI) as a function of stellar mass between cluster observables and the formation time of their dark matter halos. Upper left panel: MI between the magnitude gap (defined as the r-band magnitude difference) between the central galaxy of the group and the brightest satellite, as well as the fourth brightest satellite, and the formation time of their dark matter halos. Upper right panel: MI between the distance from the central galaxy of the group to the closest satellite, as well as the fourth closest satellite, and the formation time of their dark matter halos. Lower left panel: MI between the distance from the central galaxy of the group to the brightest satellite, as well as the fourth brightest satellite, and the formation time of their dark matter halos. Lower right panel: MI between the richness (number of satellites per halo) and the formation time of their dark matter halos.}
    
    \label{fig:Groups}
\end{figure*}

\begin{figure*}
    \centering
    \includegraphics[width=1.8\columnwidth]{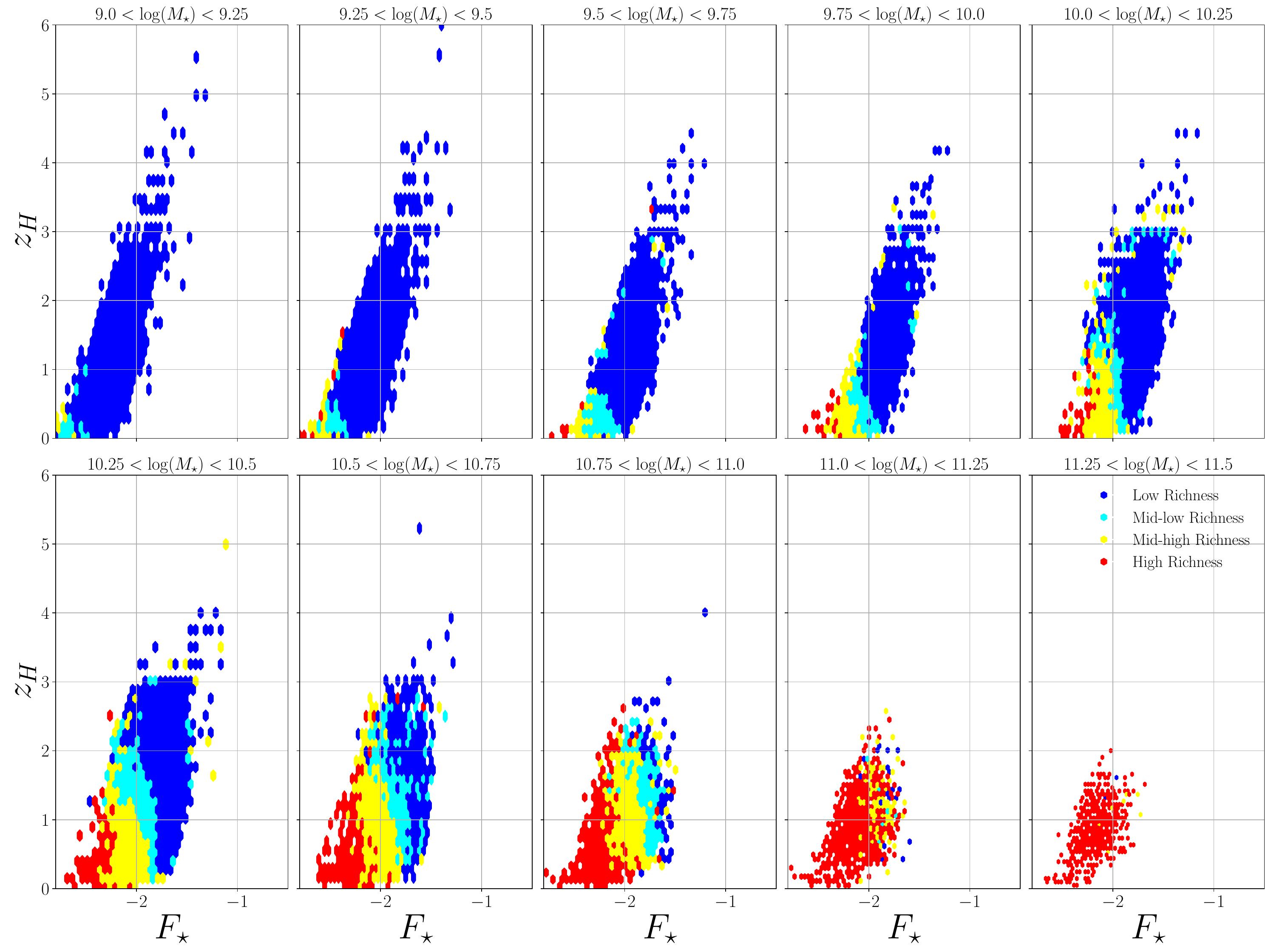}\\
    \caption{Hexbin plot of central galaxies for $z_H$ as a function of $F_\star$, colour-coding by richness quartiles.}
    \label{fig:2DRichness}
\end{figure*}

In addition to exploring the individual properties of central and satellite galaxies, we extended our analysis to examine the MI between $z_H$ and several group properties. These properties reflect the collective characteristics of galaxies within halos hosting central galaxies with stellar masses between $10^{9}$ and $10^{11.5}$ \Msunh, alongside satellite galaxies with r-band $M_r < -19$. By incorporating central satellite properties, we gain additional information on the connection between halo assembly and observable quantities, thus enhancing our understanding of potential halo age indicators.

The magnitude gap, defined as the difference in r-band magnitudes\footnote{Magnitudes are based on the summed-up luminosities of all the stellar particles in the group.} between the central galaxy and the nth brightest satellite galaxy, is widely used as a proxy for halo age and concentration. Earlier halo formation dynamically increases the likelihood that the central galaxy accretes or cannibalizes nearby massive satellites, leaving behind smaller and less luminous remnants, resulting in a larger magnitude gap \citep{2020MNRAS.493.1361F}.

\citet{2020MNRAS.493.1361F} employed haloes of mass greater than $10^{14}\, M_\odot$ from the IllustrisTNG project (TNG300 simulation). They found a strong correlation between the magnitude gap - quantified as the difference in r-band magnitude between the central galaxy and the halo's fourth brightest galaxy, with satellite galaxies limited to $M_r < -19$ -  and the formation epoch of the host halo. Similarly, \citet{2018MNRAS.474..866V} analyzed ensembles of galaxy systems divided by redshift ($0.2 < z < 0.4$) and the observed magnitude gap between the central galaxy and its brightest satellite as an indicator of formation time. They showed that systems characterized by large magnitude gaps tend to exhibit higher concentrations and are associated with earlier formation times. These results were supported by measurements of weak-lensing mass concentration and NFW parametric mass profiles derived from the SDSS Coadd dataset using the redMaPPer algorithm. The findings of both studies are consistent with the notion that the magnitude gap serves as a reliable observational indicator of halo formation time.\\

To quantify the correlation between $z_H$ and various cluster observables, we compute the MI for several measurable quantities: the r-band magnitude gap between the central galaxy and both the brightest and fourth brightest satellites (upper left panel, Figure \ref{fig:Groups}); the distance between the central galaxy and the closest satellite, as well as the fourth closest satellite (upper right panel); the distance from the central galaxy to the brightest satellite and the fourth brightest satellite (lower left panel); and the richness, defined as the number of satellites per halo with $M_r < -19$ (lower right panel). Similarly to \citet{2020MNRAS.493.1361F}, we select satellite galaxies with r-band magnitudes $M_r < -19$. However, unlike their work, our analysis includes only DM haloes containing central galaxies that span the same stellar mass described above. This methodology capitalises on the observable nature of magnitude gaps, which are frequently employed in cluster studies as proxies for halo formation history.


The results presented in Figure \ref{fig:Groups} reveal distinct patterns in the properties of satellite galaxies and their distribution within the haloes. For instance, the MI between the magnitude gaps (both for the brightest and fourth brightest satellites) and $z_H$ shows a decreasing trend as stellar mass increases across different stellar mass bins. Interestingly, MI values are slightly higher for the magnitude gap calculated using the fourth brightest satellite in massive haloes, albeit with larger uncertainties in the intermediate mass bin. This suggests that the fourth brightest satellite may serve as a better proxy for the formation history of these haloes. However, the generally weak correlations observed for both magnitude gap metrics indicate that, while related to $z_H$, these observables have limited predictive power for halo formation time. Similarly, the MI between the distance to the nearest satellite and $z_H$ is relatively high for lower stellar mass galaxies, reflecting the gravitational influence of proximity, but it decreases in more massive haloes. For the fourth closest satellite, MI values are slightly higher for intermediate stellar masses, though accompanied by larger uncertainties. In contrast, the MI for the distance to the brightest satellite remains low, suggesting a weaker correlation with halo formation history. In contrast, the MI between richness and $z_H$ consistently increases with stellar mass, indicating that richness is more closely linked to the formation history in more massive haloes. Nonetheless, the overall correlations remain weak, highlighting the complexity of the relationship between satellite properties and halo formation history.\\

Figure \ref{fig:2DRichness} extends these trends, highlighting distinct relationships between the stellar mass of central galaxies, the richness of satellite populations, and the halo formation time. For higher stellar masses ($M_\star \gtrsim 10^{10.25}$ \Msunh), central galaxies hosted by halos with low assembly time show high $z_G$. These halos exhibit low $F_\star$ and increased richness, reflecting prolonged halo growth through the accretion of satellite galaxies. In contrast, for lower stellar masses (up to the characteristic stellar mass), halos with higher $F_\star$ (and thus earlier formation histories) tend to host fewer satellites, whereas halos with lower $F_\star$ and later assembly times are associated with an increased number of satellites, supporting the idea of extended halo growth through accretion. These trends align with the above discussion, where we observed that halos with $M_h \lesssim 10^{12}$ \Msunh efficiently form stars early, but more massive halos with $M_h \gtrsim 1.44 \times 10^{12}$ \Msunh continue to grow via hierarchical mergers, accreting a larger satellite population. These results reinforce the findings by \citet{2020MNRAS.499.4748M}, which suggest that in low-mass halos, in situ star formation dominates, while in massive halos, ex situ contributions become increasingly significant, driving the observed trends in richness and $F_\star$.

The findings from Figures \ref{fig:ZFor}, \ref{fig:Groups}, and \ref{fig:2DRichness} collectively highlight the intricate interplay between halo assembly, satellite properties, and galaxy formation efficiency. In low-mass halos, central galaxies appear closely tied to their halo's formation history, as evidenced by the strong correlation between $z_G$, $z_H$, and $F_\star$. In massive halos, however, late-stage accretion of satellites, leads to higher richness, lower $F_\star$, and weaker correlations with $z_G$ of central galaxies. These observations underscore the distinct evolutionary paths of halos and galaxies, shaped by hierarchical assembly, environmental interactions, and feedback processes.

\section{Summary and conclusions}
\label{sec:conclu}

This study employed Mutual Information (MI) analysis to explore the complex interplay between galaxy properties and the assembly history of their host DM haloes, using the Illustris-TNG numerical simulation. By analyzing assembly times, specific star formation rates (sSFR), colour ($g - i$), and galaxy formation efficiency ($F_\star$), we have gained new insight into the factors shaping galaxy evolution. Our key findings can be summarized as follows.

\begin{enumerate}
    \item We find a strong correlation between the assembly time of central galaxies and their host haloes, particularly for lower-mass systems. This suggests a co-evolutionary relationship, influenced by environmental factors and feedback processes. Conversely, higher-mass central galaxies exhibit a weaker correlation, indicating that their formation history involve more complex process, such as major mergers and AGN feedback. These galaxies exhibit earlier formation times ($z_G$) compared to their host DM haloes, accompanied by redder colours $(g-i)$ and lower sSFR, consistent with the \emph{downsizing effect}, in concordance with \citet{2020MNRAS.499.4748M} who found that in massive haloes, these galaxies can only growth efficiently at high $z$. This highlights the importance of considering $F_\star$ as a more robust proxy for halo assembly time in central galaxies compared to traditional measures like colour $(g-i)$ and $sSFR$. The strong correlation between $F_\star$ and halo assembly times (with MI values reaching $0.38 - 0.43 \, nats$ and $r_{x,y}$ reaching $0.8 - 0.6$) suggests that $z_H$ plays an important role in determining some properties, such as  $F_\star$. This result is aligned with the findings of \citet{Lim_2016}. The observed correlation may be primarily attributed to increased gas loss due to supernova feedback, a phenomenon more pronounced in lower-mass systems than in their more massive counterparts. The shallow potential wells of low-mass galaxies make them more susceptible to gas outflows driven by supernova explosions and radiation pressure from massive stars, limiting their gas supply for star formation and consequently affecting their growth \citep{1986ApJ...303...39D, 2008MNRAS.387..577O,2015MNRAS.450..606L}.

    \item The MI analysis revealed a subtle yet discernible relationship between the colour $(g-i)$ and the assembly time of their host halo for central galaxies with stellar masses of up to $10^{10.5}$ \Msunh, with MI values reaching $\sim 0.18 \, nats$. For massive galaxies, the correlation was even weaker. Similarly, the correlation between the sSFR and the assembly time of their host haloes was negligible across all stellar mass bins. These results suggest that the shared information between these observables and the assembly time of their host DM haloes is limited. Factors, such as star formation, cold gas availability, mergers, and environmental effects, likely contribute more significant to the sSFR and color $(g-i)$. This lack of correlations between this observational variables with the assembly assembly time of their host DM haloes, reflects the multiple mechanisms that influence the galaxy evolution beyond the halo assembly history that come into play. 

       
    \item For satellite galaxies, the MI analysis yielded values close to zero for all evaluated mass ranges and variables, indicating that their formation and evolution are largely decoupled from the assembly history of their host dark matter haloes. Environmental factors, such as tidal stripping, ram pressure stripping, and harassment, can disrupt the correlation between the assembly time of a host halo and its satellite galaxy population. These processes prevent satellite galaxies from retain gas and dust, further limiting star formation and evolutionary growth. The pronounced downsizing effect observed in satellite galaxies markedly distinct from the behaviour of central galaxies further underscores the diverging evolutionary paths of these two galaxy populations. This distinct path is likely linked to the accretion of satellites into larger haloes, where environmental quenching mechanism play a dominant role.

    \item Our analysis of MI between cluster observables and the assembly time of their host haloes reveals that certain group properties provide insights into halo formation history, albeit with varying levels of predictive power. The magnitude gap, particularly when defined using the fourth brightest satellite, exhibits a weak correlation with halo assembly time, suggesting its limited utility as a stand-alone indicator. Similarly, MI values for the distances between the central galaxy and its nearest or fourth nearest satellites show modest correlations for low-mass haloes but diminish at higher stellar masses, highlighting the influence of gravitational interactions and environmental processes in shaping these distributions. The richness of satellite populations emerges as a slightly stronger proxy, with MI values increasing consistently with stellar mass. This trend suggest that in massive haloes, richness correlates more closely with halo assembly history than other metrics, such as the magnitude gap, the distance to the closest satellite, or the distance to the brightest satellite. This reflects prolonged hierarchical growth and satellite accretion in such systems. Conversely, for low-mass haloes, high $F_\star$ and early assembly times are associated with lower richness, while low $F_\star$ and late assembly times correspond to an increase in the number of satellites. These findings align with the trends observed in Figures \ref{fig:Groups} and \ref{fig:2DRichness}, which illustrate the interplay between satellite properties, halo growth, and galaxy formation efficiency. However, the overall weak correlations across most group observables emphasise the complexity of the relationship between halo assembly history and satellite populations. The results suggest that environmental processes, such as tidal stripping and mergers, significantly influence satellite properties, diminishing the direct imprint of halo formation time. A more nuanced understanding may require combining multiple observables to unravel the intricate co-evolution of central and satellite galaxies within their host haloes.
\end{enumerate}

The strong correlation observed between the assembly times of central galaxies, $F_\star$, and the assembly times of their host DM haloes underscores a co-evolutionary relationship driven by feedback processes and environmental factors. Less massive galaxies, with their shallower potential wells and lower gas fractions, appear more vulnerable to supernova feedback, which may shape their evolutionary paths and strengthen their connection to halo assembly time. In contrast, the weaker correlations observed for colour $(g-i)$, sSFR, and satellite galaxies highlight the multifaceted nature of galaxy formation. These results emphasize the necessity of considering both internal mechanisms, such as feedback, and external processes, such as environmental interactions, to fully understand the evolution of the galaxy. Future studies could explore the role of CGM and other environmental factors in greater detail to elucidate the precise mechanisms driving these observed trends.

\section*{Acknowledgements}
We thank Forero-Romero J. E. for useful discussions. We are very grateful to our referee for the careful reading of the paper and for his/her comments and detailed suggestions.

\section*{Data availability}
The datasets were derived from sources in the public domain \url{https://www.tng-project.org/data/}.

\bibliographystyle{mnras}
\bibliography{bibliography}

\begin{thebibliography}{}
\makeatletter
\relax
\def\mn@urlcharsother{\let\do\@makeother \do\$\do\&\do\#\do\^\do\_\do\%\do\~}
\def\mn@doi{\begingroup\mn@urlcharsother \@ifnextchar [ {\mn@doi@} {\mn@doi@[]}}
\def\mn@doi@[#1]#2{\def\@tempa{#1}\ifx\@tempa\@empty \href {http://dx.doi.org/#2} {doi:#2}\else \href {http://dx.doi.org/#2} {#1}\fi \endgroup}
\def\mn@eprint#1#2{\mn@eprint@#1:#2::\@nil}
\def\mn@eprint@arXiv#1{\href {http://arxiv.org/abs/#1} {{\tt arXiv:#1}}}
\def\mn@eprint@dblp#1{\href {http://dblp.uni-trier.de/rec/bibtex/#1.xml} {dblp:#1}}
\def\mn@eprint@#1:#2:#3:#4\@nil{\def\@tempa {#1}\def\@tempb {#2}\def\@tempc {#3}\ifx \@tempc \@empty \let \@tempc \@tempb \let \@tempb \@tempa \fi \ifx \@tempb \@empty \def\@tempb {arXiv}\fi \@ifundefined {mn@eprint@\@tempb}{\@tempb:\@tempc}{\expandafter \expandafter \csname mn@eprint@\@tempb\endcsname \expandafter{\@tempc}}}

\bibitem[\protect\citeauthoryear{{Artale}, {Zehavi}, {Contreras}  \& {Norberg}}{{Artale} et~al.}{2018}]{2018MNRAS.480.3978A}
{Artale} M.~C.,  {Zehavi} I.,  {Contreras} S.,   {Norberg} P.,  2018, \mn@doi [\mnras] {10.1093/mnras/sty2110}, \href {https://ui.adsabs.harvard.edu/abs/2018MNRAS.480.3978A} {480, 3978}

\bibitem[\protect\citeauthoryear{{Behroozi}, {Wechsler}, {Hearin}  \& {Conroy}}{{Behroozi} et~al.}{2019}]{2019MNRAS.488.3143B}
{Behroozi} P.,  {Wechsler} R.~H.,  {Hearin} A.~P.,   {Conroy} C.,  2019, \mn@doi [\mnras] {10.1093/mnras/stz1182}, \href {https://ui.adsabs.harvard.edu/abs/2019MNRAS.488.3143B} {488, 3143}

\bibitem[\protect\citeauthoryear{{Berti}, {Dawson}  \& {Dominguez}}{{Berti} et~al.}{2023}]{Berti_2023}
{Berti} A.~M.,  {Dawson} K.~S.,   {Dominguez} W.,  2023, \mn@doi [arXiv e-prints] {10.48550/arXiv.2303.16096}, \href {https://ui.adsabs.harvard.edu/abs/2023arXiv230316096B} {p. arXiv:2303.16096}

\bibitem[\protect\citeauthoryear{{Bhattacharjee}, {Pandey}  \& {Sarkar}}{{Bhattacharjee} et~al.}{2020}]{2020JCAP...09..039B}
{Bhattacharjee} S.,  {Pandey} B.,   {Sarkar} S.,  2020, \mn@doi [\jcap] {10.1088/1475-7516/2020/09/039}, \href {https://ui.adsabs.harvard.edu/abs/2020JCAP...09..039B} {2020, 039}

\bibitem[\protect\citeauthoryear{{Bose}, {Eisenstein}, {Hernquist}, {Pillepich}, {Nelson}, {Marinacci}, {Springel}  \& {Vogelsberger}}{{Bose} et~al.}{2019}]{2019MNRAS.490.5693B}
{Bose} S.,  {Eisenstein} D.~J.,  {Hernquist} L.,  {Pillepich} A.,  {Nelson} D.,  {Marinacci} F.,  {Springel} V.,   {Vogelsberger} M.,  2019, \mn@doi [\mnras] {10.1093/mnras/stz2546}, \href {https://ui.adsabs.harvard.edu/abs/2019MNRAS.490.5693B} {490, 5693}

\bibitem[\protect\citeauthoryear{Bower, Schaye, Frenk, Theuns, Schaller, Crain  \& McAlpine}{Bower et~al.}{2016}]{10.1093/mnras/stw2735}
Bower R.~G.,  Schaye J.,  Frenk C.~S.,  Theuns T.,  Schaller M.,  Crain R.~A.,   McAlpine S.,  2016, \mn@doi [Monthly Notices of the Royal Astronomical Society] {10.1093/mnras/stw2735}, 465, 32

\bibitem[\protect\citeauthoryear{{Busch} \& {White}}{{Busch} \& {White}}{2017}]{2017MNRAS.470.4767B}
{Busch} P.,  {White} S. D.~M.,  2017, \mn@doi [\mnras] {10.1093/mnras/stx1584}, \href {https://ui.adsabs.harvard.edu/abs/2017MNRAS.470.4767B} {470, 4767}

\bibitem[\protect\citeauthoryear{{Contreras}, {Zehavi}, {Padilla}, {Baugh}, {Jim{\'e}nez}  \& {Lacerna}}{{Contreras} et~al.}{2019}]{2019MNRAS.484.1133C}
{Contreras} S.,  {Zehavi} I.,  {Padilla} N.,  {Baugh} C.~M.,  {Jim{\'e}nez} E.,   {Lacerna} I.,  2019, \mn@doi [\mnras] {10.1093/mnras/stz018}, \href {https://ui.adsabs.harvard.edu/abs/2019MNRAS.484.1133C} {484, 1133}

\bibitem[\protect\citeauthoryear{{Croton}, {Gao}  \& {White}}{{Croton} et~al.}{2007}]{2007MNRAS.374.1303C}
{Croton} D.~J.,  {Gao} L.,   {White} S.~D.~M.,  2007, \mn@doi [mnras] {10.1111/j.1365-2966.2006.11230.x}, \href {http://adsabs.harvard.edu/abs/2007MNRAS.374.1303C} {374, 1303}

\bibitem[\protect\citeauthoryear{{Cui}, {Dav{\'e}}, {Peacock}, {Angl{\'e}s-Alc{\'a}zar}  \& {Yang}}{{Cui} et~al.}{2021}]{2021NatAs...5.1069C}
{Cui} W.,  {Dav{\'e}} R.,  {Peacock} J.~A.,  {Angl{\'e}s-Alc{\'a}zar} D.,   {Yang} X.,  2021, \mn@doi [Nature Astronomy] {10.1038/s41550-021-01404-1}, \href {https://ui.adsabs.harvard.edu/abs/2021NatAs...5.1069C} {5, 1069}

\bibitem[\protect\citeauthoryear{{Dekel} \& {Silk}}{{Dekel} \& {Silk}}{1986}]{1986ApJ...303...39D}
{Dekel} A.,  {Silk} J.,  1986, \mn@doi [\apj] {10.1086/164050}, \href {https://ui.adsabs.harvard.edu/abs/1986ApJ...303...39D} {303, 39}

\bibitem[\protect\citeauthoryear{{Dvornik} et~al.,}{{Dvornik} et~al.}{2017}]{2017MNRAS.468.3251D}
{Dvornik} A.,  et~al., 2017, \mn@doi [\mnras] {10.1093/mnras/stx705}, \href {https://ui.adsabs.harvard.edu/abs/2017MNRAS.468.3251D} {468, 3251}

\bibitem[\protect\citeauthoryear{{Farahi}, {Ho}  \& {Trac}}{{Farahi} et~al.}{2020}]{2020MNRAS.493.1361F}
{Farahi} A.,  {Ho} M.,   {Trac} H.,  2020, \mn@doi [\mnras] {10.1093/mnras/staa291}, \href {https://ui.adsabs.harvard.edu/abs/2020MNRAS.493.1361F} {493, 1361}

\bibitem[\protect\citeauthoryear{{Gao}, {Springel}  \& {White}}{{Gao} et~al.}{2005}]{2005MNRAS.363L..66G}
{Gao} L.,  {Springel} V.,   {White} S. D.~M.,  2005, \mn@doi [\mnras] {10.1111/j.1745-3933.2005.00084.x}, \href {https://ui.adsabs.harvard.edu/abs/2005MNRAS.363L..66G} {363, L66}

\bibitem[\protect\citeauthoryear{{Gu}, {Conroy}  \& {Behroozi}}{{Gu} et~al.}{2016}]{2016ApJ...833....2G}
{Gu} M.,  {Conroy} C.,   {Behroozi} P.,  2016, \mn@doi [\apj] {10.3847/0004-637X/833/1/2}, \href {https://ui.adsabs.harvard.edu/abs/2016ApJ...833....2G} {833, 2}

\bibitem[\protect\citeauthoryear{Hearin \& Watson}{Hearin \& Watson}{2013}]{10.1093/mnras/stt1374}
Hearin A.~P.,  Watson D.~F.,  2013, Monthly Notices of the Royal Astronomical Society, 435, 1313

\bibitem[\protect\citeauthoryear{{Hearin}, {Watson}, {Becker}, {Reyes}, {Berlind}  \& {Zentner}}{{Hearin} et~al.}{2014}]{2014MNRAS.444..729H}
{Hearin} A.~P.,  {Watson} D.~F.,  {Becker} M.~R.,  {Reyes} R.,  {Berlind} A.~A.,   {Zentner} A.~R.,  2014, \mn@doi [\mnras] {10.1093/mnras/stu1443}, \href {https://ui.adsabs.harvard.edu/abs/2014MNRAS.444..729H} {444, 729}

\bibitem[\protect\citeauthoryear{{Jackson}, {Pasquali}, {Pacifici}, {Engler}, {Pillepich}  \& {Grebel}}{{Jackson} et~al.}{2020}]{2020MNRAS.497.4262J}
{Jackson} T.~M.,  {Pasquali} A.,  {Pacifici} C.,  {Engler} C.,  {Pillepich} A.,   {Grebel} E.~K.,  2020, \mn@doi [\mnras] {10.1093/mnras/staa2306}, \href {https://ui.adsabs.harvard.edu/abs/2020MNRAS.497.4262J} {497, 4262}

\bibitem[\protect\citeauthoryear{{Jung}, {Lee}  \& {Yi}}{{Jung} et~al.}{2014}]{2014ApJ...794...74J}
{Jung} I.,  {Lee} J.,   {Yi} S.~K.,  2014, \mn@doi [\apj] {10.1088/0004-637X/794/1/74}, \href {https://ui.adsabs.harvard.edu/abs/2014ApJ...794...74J} {794, 74}

\bibitem[\protect\citeauthoryear{Kinney \& Atwal}{Kinney \& Atwal}{2014}]{pnas.1309933111}
Kinney J.~B.,  Atwal G.~S.,  2014, \mn@doi [Proceedings of the National Academy of Sciences] {10.1073/pnas.1309933111}, 111, 3354

\bibitem[\protect\citeauthoryear{{Lacerna}, {Padilla}  \& {Stasyszyn}}{{Lacerna} et~al.}{2014}]{Lacerna_2014}
{Lacerna} I.,  {Padilla} N.,   {Stasyszyn} F.,  2014, \mn@doi [\mnras] {10.1093/mnras/stu1318}, \href {https://ui.adsabs.harvard.edu/abs/2014MNRAS.443.3107L} {443, 3107}

\bibitem[\protect\citeauthoryear{{Li}, {Mo}  \& {Gao}}{{Li} et~al.}{2008}]{2008MNRAS.389.1419L}
{Li} Y.,  {Mo} H.~J.,   {Gao} L.,  2008, \mn@doi [\mnras] {10.1111/j.1365-2966.2008.13667.x}, \href {https://ui.adsabs.harvard.edu/abs/2008MNRAS.389.1419L} {389, 1419}

\bibitem[\protect\citeauthoryear{Lim, Mo, Wang  \& Yang}{Lim et~al.}{2016}]{Lim_2016}
Lim S.~H.,  Mo H.,  Wang H.,   Yang X.,  2016, \mn@doi [Monthly Notices of the Royal Astronomical Society] {10.1093/mnras/stv2282}

\bibitem[\protect\citeauthoryear{{Lin}, {Mandelbaum}, {Huang}, {Huang}, {Dalal}, {Diemer}, {Jian}  \& {Kravtsov}}{{Lin} et~al.}{2016}]{2016ApJ...819..119L}
{Lin} Y.-T.,  {Mandelbaum} R.,  {Huang} Y.-H.,  {Huang} H.-J.,  {Dalal} N.,  {Diemer} B.,  {Jian} H.-Y.,   {Kravtsov} A.,  2016, \mn@doi [\apj] {10.3847/0004-637X/819/2/119}, \href {https://ui.adsabs.harvard.edu/abs/2016ApJ...819..119L} {819, 119}

\bibitem[\protect\citeauthoryear{{Lu}, {Mo}  \& {Lu}}{{Lu} et~al.}{2015}]{2015MNRAS.450..606L}
{Lu} Z.,  {Mo} H.~J.,   {Lu} Y.,  2015, \mn@doi [\mnras] {10.1093/mnras/stv671}, \href {https://ui.adsabs.harvard.edu/abs/2015MNRAS.450..606L} {450, 606}

\bibitem[\protect\citeauthoryear{{Lucie-Smith}, {Peiris}, {Pontzen}, {Nord}, {Thiyagalingam}  \& {Piras}}{{Lucie-Smith} et~al.}{2022}]{2022PhRvD.105j3533L}
{Lucie-Smith} L.,  {Peiris} H.~V.,  {Pontzen} A.,  {Nord} B.,  {Thiyagalingam} J.,   {Piras} D.,  2022, \mn@doi [\prd] {10.1103/PhysRevD.105.103533}, \href {https://ui.adsabs.harvard.edu/abs/2022PhRvD.105j3533L} {105, 103533}

\bibitem[\protect\citeauthoryear{{Martizzi}, {Vogelsberger}, {Torrey}, {Pillepich}, {Hansen}, {Marinacci}  \& {Hernquist}}{{Martizzi} et~al.}{2020}]{2020MNRAS.491.5747M}
{Martizzi} D.,  {Vogelsberger} M.,  {Torrey} P.,  {Pillepich} A.,  {Hansen} S.~H.,  {Marinacci} F.,   {Hernquist} L.,  2020, \mn@doi [\mnras] {10.1093/mnras/stz3418}, \href {https://ui.adsabs.harvard.edu/abs/2020MNRAS.491.5747M} {491, 5747}

\bibitem[\protect\citeauthoryear{{Matthee}, {Schaye}, {Crain}, {Schaller}, {Bower}  \& {Theuns}}{{Matthee} et~al.}{2017}]{2017MNRAS.465.2381M}
{Matthee} J.,  {Schaye} J.,  {Crain} R.~A.,  {Schaller} M.,  {Bower} R.,   {Theuns} T.,  2017, \mn@doi [\mnras] {10.1093/mnras/stw2884}, \href {https://ui.adsabs.harvard.edu/abs/2017MNRAS.465.2381M} {465, 2381}

\bibitem[\protect\citeauthoryear{Mezard \& Montanari}{Mezard \& Montanari}{2009}]{10.5555/1592967}
Mezard M.,  Montanari A.,  2009, Information, Physics, and Computation.
Oxford University Press, Inc., USA

\bibitem[\protect\citeauthoryear{{Miyatake}, {More}, {Takada}, {Spergel}, {Mandelbaum}, {Rykoff}  \& {Rozo}}{{Miyatake} et~al.}{2016}]{2016PhRvL.116d1301M}
{Miyatake} H.,  {More} S.,  {Takada} M.,  {Spergel} D.~N.,  {Mandelbaum} R.,  {Rykoff} E.~S.,   {Rozo} E.,  2016, \mn@doi [Physical Review Letters] {10.1103/PhysRevLett.116.041301}, \href {http://adsabs.harvard.edu/abs/2016PhRvL.116d1301M} {116, 041301}

\bibitem[\protect\citeauthoryear{{Montero-Dorta} et~al.,}{{Montero-Dorta} et~al.}{2020}]{2020MNRAS.tmp.1844M}
{Montero-Dorta} A.~D.,  et~al., 2020, \mn@doi [\mnras] {10.1093/mnras/staa1624}, \href {https://ui.adsabs.harvard.edu/abs/2020MNRAS.tmp.1844M} {}

\bibitem[\protect\citeauthoryear{{Montero-Dorta}, {Chaves-Montero}, {Artale}  \& {Favole}}{{Montero-Dorta} et~al.}{2021}]{2021MNRAS.508..940M}
{Montero-Dorta} A.~D.,  {Chaves-Montero} J.,  {Artale} M.~C.,   {Favole} G.,  2021, \mn@doi [\mnras] {10.1093/mnras/stab2556}, \href {https://ui.adsabs.harvard.edu/abs/2021MNRAS.508..940M} {508, 940}

\bibitem[\protect\citeauthoryear{{Moster}, {Naab}  \& {White}}{{Moster} et~al.}{2020}]{2020MNRAS.499.4748M}
{Moster} B.~P.,  {Naab} T.,   {White} S. D.~M.,  2020, \mn@doi [\mnras] {10.1093/mnras/staa3019}, \href {https://ui.adsabs.harvard.edu/abs/2020MNRAS.499.4748M} {499, 4748}

\bibitem[\protect\citeauthoryear{Naree \& Muanwong}{Naree \& Muanwong}{2023}]{Naree_2023}
Naree N.,  Muanwong O.,  2023, \mn@doi [Journal of Physics: Conference Series] {10.1088/1742-6596/2431/1/012083}, 2431, 012083

\bibitem[\protect\citeauthoryear{{Nelson} et~al.,}{{Nelson} et~al.}{2019}]{2019ComAC...6....2N}
{Nelson} D.,  et~al., 2019, \mn@doi [Computational Astrophysics and Cosmology] {10.1186/s40668-019-0028-x}, \href {https://ui.adsabs.harvard.edu/abs/2019ComAC...6....2N} {6, 2}

\bibitem[\protect\citeauthoryear{{Niemiec} et~al.,}{{Niemiec} et~al.}{2018}]{2018MNRAS.477L...1N}
{Niemiec} A.,  et~al., 2018, \mn@doi [\mnras] {10.1093/mnrasl/sly041}, \href {https://ui.adsabs.harvard.edu/abs/2018MNRAS.477L...1N} {477, L1}

\bibitem[\protect\citeauthoryear{{Niemiec}, {Giocoli}, {Cohen}, {Jauzac}, {Jullo}  \& {Limousin}}{{Niemiec} et~al.}{2022}]{2022MNRAS.tmp..806N}
{Niemiec} A.,  {Giocoli} C.,  {Cohen} E.,  {Jauzac} M.,  {Jullo} E.,   {Limousin} M.,  2022, \mn@doi [\mnras] {10.1093/mnras/stac832}, \href {https://ui.adsabs.harvard.edu/abs/2022MNRAS.tmp..806N} {}

\bibitem[\protect\citeauthoryear{{Oppenheimer} \& {Dav{\'e}}}{{Oppenheimer} \& {Dav{\'e}}}{2008}]{2008MNRAS.387..577O}
{Oppenheimer} B.~D.,  {Dav{\'e}} R.,  2008, \mn@doi [\mnras] {10.1111/j.1365-2966.2008.13280.x}, \href {https://ui.adsabs.harvard.edu/abs/2008MNRAS.387..577O} {387, 577}

\bibitem[\protect\citeauthoryear{{Pandey}}{{Pandey}}{2017}]{2017MNRAS.471L..77P}
{Pandey} B.,  2017, \mn@doi [\mnras] {10.1093/mnrasl/slx109}, \href {https://ui.adsabs.harvard.edu/abs/2017MNRAS.471L..77P} {471, L77}

\bibitem[\protect\citeauthoryear{Piras, Peiris, Pontzen, Lucie-Smith, Guo  \& Nord}{Piras et~al.}{2023}]{Piras23}
Piras D.,  Peiris H.~V.,  Pontzen A.,  Lucie-Smith L.,  Guo N.,   Nord B.,  2023, \mn@doi [Machine Learning: Science and Technology] {10.1088/2632-2153/acc444}, 4, 025006

\bibitem[\protect\citeauthoryear{{Planck Collaboration} et~al.,}{{Planck Collaboration} et~al.}{2016}]{2016A&A...594A..13P}
{Planck Collaboration} et~al., 2016, \mn@doi [\aap] {10.1051/0004-6361/201525830}, \href {https://ui.adsabs.harvard.edu/abs/2016A&A...594A..13P} {594, A13}

\bibitem[\protect\citeauthoryear{{Rodriguez-Gomez} et~al.,}{{Rodriguez-Gomez} et~al.}{2015}]{2015MNRAS.449...49R}
{Rodriguez-Gomez} V.,  et~al., 2015, \mn@doi [\mnras] {10.1093/mnras/stv264}, \href {https://ui.adsabs.harvard.edu/abs/2015MNRAS.449...49R} {449, 49}

\bibitem[\protect\citeauthoryear{{Rodr{\'\i}guez Montero}, {Dav{\'e}}, {Wild}, {Angl{\'e}s-Alc{\'a}zar}  \& {Narayanan}}{{Rodr{\'\i}guez Montero} et~al.}{2019}]{2019MNRAS.490.2139R}
{Rodr{\'\i}guez Montero} F.,  {Dav{\'e}} R.,  {Wild} V.,  {Angl{\'e}s-Alc{\'a}zar} D.,   {Narayanan} D.,  2019, \mn@doi [\mnras] {10.1093/mnras/stz2580}, \href {https://ui.adsabs.harvard.edu/abs/2019MNRAS.490.2139R} {490, 2139}

\bibitem[\protect\citeauthoryear{Salcedo, Maller, Berlind, Sinha, McBride, Behroozi, Wechsler  \& Weinberg}{Salcedo et~al.}{2018}]{10.1093/mnras/sty109}
Salcedo A.~N.,  Maller A.~H.,  Berlind A.~A.,  Sinha M.,  McBride C.~K.,  Behroozi P.~S.,  Wechsler R.~H.,   Weinberg D.~H.,  2018, Monthly Notices of the Royal Astronomical Society, 475, 4411

\bibitem[\protect\citeauthoryear{{Sarkar} \& {Pandey}}{{Sarkar} \& {Pandey}}{2020}]{2020MNRAS.497.4077S}
{Sarkar} S.,  {Pandey} B.,  2020, \mn@doi [\mnras] {10.1093/mnras/staa2236}, \href {https://ui.adsabs.harvard.edu/abs/2020MNRAS.497.4077S} {497, 4077}

\bibitem[\protect\citeauthoryear{Shannon}{Shannon}{1948}]{shannon1948mathematical}
Shannon C.~E.,  1948, \mn@doi [Bell System Technical Journal] {https://doi.org/10.1002/j.1538-7305.1948.tb01338.x}, 27, 379

\bibitem[\protect\citeauthoryear{{Springel}}{{Springel}}{2010}]{2010MNRAS.401..791S}
{Springel} V.,  2010, \mn@doi [\mnras] {10.1111/j.1365-2966.2009.15715.x}, \href {https://ui.adsabs.harvard.edu/abs/2010MNRAS.401..791S} {401, 791}

\bibitem[\protect\citeauthoryear{{Springel} et~al.,}{{Springel} et~al.}{2018}]{2018MNRAS.475..676S}
{Springel} V.,  et~al., 2018, \mn@doi [mnras] {10.1093/mnras/stx3304}, \href {http://adsabs.harvard.edu/abs/2018MNRAS.475..676S} {475, 676}

\bibitem[\protect\citeauthoryear{{Sui}, {Zhao}, {Jing}  \& {Mao}}{{Sui} et~al.}{2023}]{2023arXiv230704994S}
{Sui} C.,  {Zhao} X.,  {Jing} T.,   {Mao} Y.,  2023, \mn@doi [arXiv e-prints] {10.48550/arXiv.2307.04994}, \href {https://ui.adsabs.harvard.edu/abs/2023arXiv230704994S} {p. arXiv:2307.04994}

\bibitem[\protect\citeauthoryear{{Sunayama} \& {More}}{{Sunayama} \& {More}}{2019}]{2019MNRAS.490.4945S}
{Sunayama} T.,  {More} S.,  2019, \mn@doi [\mnras] {10.1093/mnras/stz2832}, \href {https://ui.adsabs.harvard.edu/abs/2019MNRAS.490.4945S} {490, 4945}

\bibitem[\protect\citeauthoryear{{Sunayama}, {More}  \& {Miyatake}}{{Sunayama} et~al.}{2022}]{2022arXiv220503277S}
{Sunayama} T.,  {More} S.,   {Miyatake} H.,  2022, arXiv e-prints, \href {https://ui.adsabs.harvard.edu/abs/2022arXiv220503277S} {p. arXiv:2205.03277}

\bibitem[\protect\citeauthoryear{{Thomas}, {Dav{\'e}}, {Angl{\'e}s-Alc{\'a}zar}  \& {Jarvis}}{{Thomas} et~al.}{2019}]{2019MNRAS.487.5764T}
{Thomas} N.,  {Dav{\'e}} R.,  {Angl{\'e}s-Alc{\'a}zar} D.,   {Jarvis} M.,  2019, \mn@doi [\mnras] {10.1093/mnras/stz1703}, \href {https://ui.adsabs.harvard.edu/abs/2019MNRAS.487.5764T} {487, 5764}

\bibitem[\protect\citeauthoryear{{Tinker}, {Wetzel}, {Conroy}  \& {Mao}}{{Tinker} et~al.}{2017}]{2017MNRAS.472.2504T}
{Tinker} J.~L.,  {Wetzel} A.~R.,  {Conroy} C.,   {Mao} Y.-Y.,  2017, \mn@doi [mnras] {10.1093/mnras/stx2066}, \href {http://adsabs.harvard.edu/abs/2017MNRAS.472.2504T} {472, 2504}

\bibitem[\protect\citeauthoryear{{Tinker}, {Hahn}, {Mao}  \& {Wetzel}}{{Tinker} et~al.}{2018}]{2018MNRAS.478.4487T}
{Tinker} J.~L.,  {Hahn} C.,  {Mao} Y.-Y.,   {Wetzel} A.~R.,  2018, \mn@doi [\mnras] {10.1093/mnras/sty1263}, \href {https://ui.adsabs.harvard.edu/abs/2018MNRAS.478.4487T} {478, 4487}

\bibitem[\protect\citeauthoryear{{Vergara} \& {Est{\'e}vez}}{{Vergara} \& {Est{\'e}vez}}{2015}]{2015arXiv150907577V}
{Vergara} J.~R.,  {Est{\'e}vez} P.~A.,  2015, \mn@doi [arXiv e-prints] {10.48550/arXiv.1509.07577}, \href {https://ui.adsabs.harvard.edu/abs/2015arXiv150907577V} {p. arXiv:1509.07577}

\bibitem[\protect\citeauthoryear{{Vitorelli}, {Cypriano}, {Makler}, {Pereira}, {Erben}  \& {Moraes}}{{Vitorelli} et~al.}{2018}]{2018MNRAS.474..866V}
{Vitorelli} A.~Z.,  {Cypriano} E.~S.,  {Makler} M.,  {Pereira} M. E.~S.,  {Erben} T.,   {Moraes} B.,  2018, \mn@doi [\mnras] {10.1093/mnras/stx2791}, \href {https://ui.adsabs.harvard.edu/abs/2018MNRAS.474..866V} {474, 866}

\bibitem[\protect\citeauthoryear{{Vogelsberger} et~al.,}{{Vogelsberger} et~al.}{2014}]{2014MNRAS.444.1518V}
{Vogelsberger} M.,  et~al., 2014, \mn@doi [mnras] {10.1093/mnras/stu1536}, \href {http://adsabs.harvard.edu/abs/2014MNRAS.444.1518V} {444, 1518}

\bibitem[\protect\citeauthoryear{{Wang}, {Yang}, {Mo}, {van den Bosch}, {Weinmann}  \& {Chu}}{{Wang} et~al.}{2008}]{2008ApJ...687..919W}
{Wang} Y.,  {Yang} X.,  {Mo} H.~J.,  {van den Bosch} F.~C.,  {Weinmann} S.~M.,   {Chu} Y.,  2008, \mn@doi [\apj] {10.1086/591836}, \href {https://ui.adsabs.harvard.edu/abs/2008ApJ...687..919W} {687, 919}

\bibitem[\protect\citeauthoryear{{Wang}, {Mo}, {Jing}, {Yang}  \& {Wang}}{{Wang} et~al.}{2011}]{2011MNRAS.413.1973W}
{Wang} H.,  {Mo} H.~J.,  {Jing} Y.~P.,  {Yang} X.,   {Wang} Y.,  2011, \mn@doi [\mnras] {10.1111/j.1365-2966.2011.18301.x}, \href {https://ui.adsabs.harvard.edu/abs/2011MNRAS.413.1973W} {413, 1973}

\bibitem[\protect\citeauthoryear{{Wechsler} \& {Tinker}}{{Wechsler} \& {Tinker}}{2018}]{2018ARA&A..56..435W}
{Wechsler} R.~H.,  {Tinker} J.~L.,  2018, \mn@doi [araa] {10.1146/annurev-astro-081817-051756}, \href {https://ui.adsabs.harvard.edu/abs/2018ARA&A..56..435W} {56, 435}

\bibitem[\protect\citeauthoryear{{Wechsler}, {Zentner}, {Bullock}, {Kravtsov}  \& {Allgood}}{{Wechsler} et~al.}{2006}]{2006ApJ...652...71W}
{Wechsler} R.~H.,  {Zentner} A.~R.,  {Bullock} J.~S.,  {Kravtsov} A.~V.,   {Allgood} B.,  2006, \mn@doi [\apj] {10.1086/507120}, \href {https://ui.adsabs.harvard.edu/abs/2006ApJ...652...71W} {652, 71}

\bibitem[\protect\citeauthoryear{{Xu} \& {Zheng}}{{Xu} \& {Zheng}}{2020}]{2020MNRAS.492.2739X}
{Xu} X.,  {Zheng} Z.,  2020, \mn@doi [\mnras] {10.1093/mnras/staa009}, \href {https://ui.adsabs.harvard.edu/abs/2020MNRAS.492.2739X} {492, 2739}

\bibitem[\protect\citeauthoryear{{Xu} et~al.,}{{Xu} et~al.}{2020}]{2020MNRAS.498.1839X}
{Xu} W.,  et~al., 2020, \mn@doi [\mnras] {10.1093/mnras/staa2497}, \href {https://ui.adsabs.harvard.edu/abs/2020MNRAS.498.1839X} {498, 1839}

\bibitem[\protect\citeauthoryear{{Yang}, {Mo}  \& {van den Bosch}}{{Yang} et~al.}{2006}]{2006ApJ...638L..55Y}
{Yang} X.,  {Mo} H.~J.,   {van den Bosch} F.~C.,  2006, \mn@doi [\apjl] {10.1086/501069}, \href {https://ui.adsabs.harvard.edu/abs/2006ApJ...638L..55Y} {638, L55}

\bibitem[\protect\citeauthoryear{{Zehavi}, {Contreras}, {Padilla}, {Smith}, {Baugh}  \& {Norberg}}{{Zehavi} et~al.}{2018}]{2018ApJ...853...84Z}
{Zehavi} I.,  {Contreras} S.,  {Padilla} N.,  {Smith} N.~J.,  {Baugh} C.~M.,   {Norberg} P.,  2018, \mn@doi [apj] {10.3847/1538-4357/aaa54a}, \href {http://adsabs.harvard.edu/abs/2018ApJ...853...84Z} {853, 84}

\bibitem[\protect\citeauthoryear{{Zentner}, {Hearin}, {van den Bosch}, {Lange}  \& {Villarreal}}{{Zentner} et~al.}{2019}]{2019MNRAS.485.1196Z}
{Zentner} A.~R.,  {Hearin} A.,  {van den Bosch} F.~C.,  {Lange} J.~U.,   {Villarreal} A.,  2019, \mn@doi [\mnras] {10.1093/mnras/stz470}, \href {https://ui.adsabs.harvard.edu/abs/2019MNRAS.485.1196Z} {485, 1196}

\bibitem[\protect\citeauthoryear{{Zu}, {Mandelbaum}, {Simet}, {Rozo}  \& {Rykoff}}{{Zu} et~al.}{2017}]{2017MNRAS.470..551Z}
{Zu} Y.,  {Mandelbaum} R.,  {Simet} M.,  {Rozo} E.,   {Rykoff} E.~S.,  2017, \mn@doi [\mnras] {10.1093/mnras/stx1264}, \href {https://ui.adsabs.harvard.edu/abs/2017MNRAS.470..551Z} {470, 551}

\makeatother
\end{thebibliography}

\appendix
\section{Influence of AGN and Merges in Galaxy Evolution}
\label{sec:ApenAGN}

\begin{figure*}
    \centering
    \includegraphics[width=1.8\columnwidth]{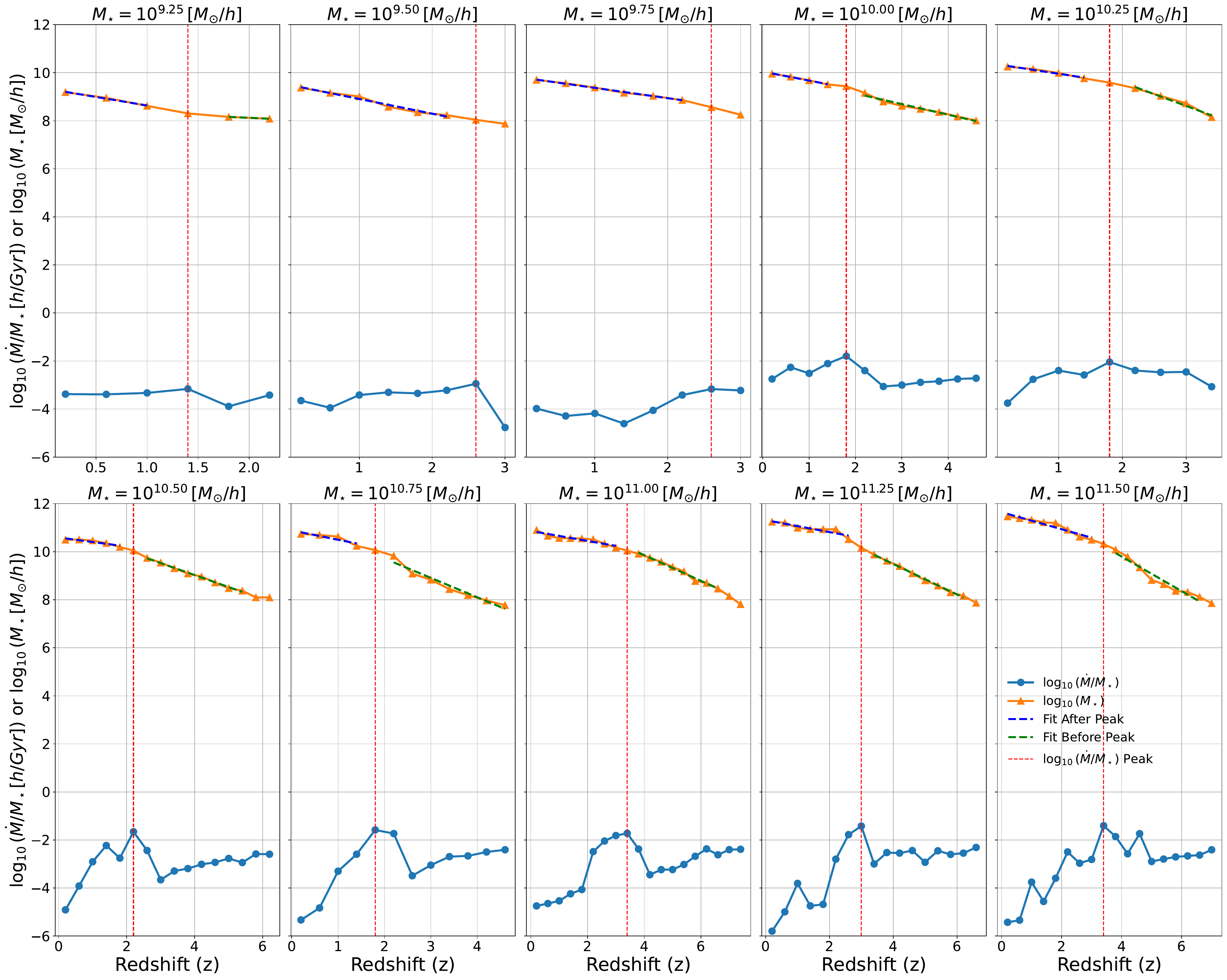}
    \caption{Evolution of the ratio $\log(\dot{M}) /\log(M_\star)$ (blue line), where $\dot{M}$ represents the sum of instantaneous accretion rates of all black holes in a subhalo, alongside the stellar mass growth of central galaxies (orange line) as a function of $z$. The red dashed line indicates the redshift corresponding to the peak in $\log(\dot{M}) /\log(M_\star)$ for each stellar mass bin on the most massive central galaxy from each stellar mass bin in our data sample.}
   \label{fig:AGNFeedback}
\end{figure*}

\begin{figure*}
    \centering
    \includegraphics[width=1.8\columnwidth]{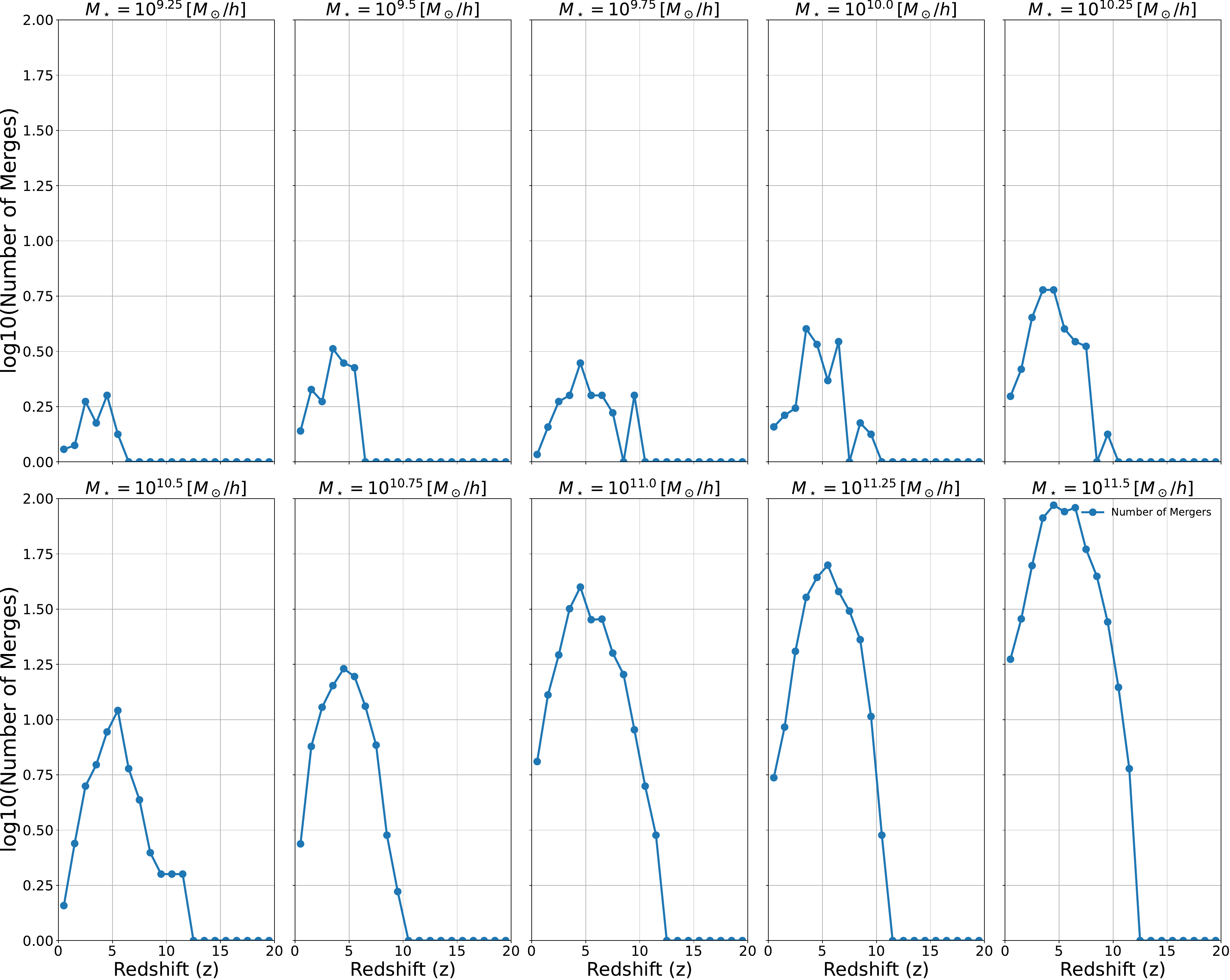}
    \caption{Number of merger events as a function of $z$ for central galaxies, divided into stellar mass bins on the most massive central galaxy from each stellar mass bin in our data sample.} 
   \label{fig:Mergers}
\end{figure*}

AGN feedback and merging processes, which primarily affect the baryonic component of galaxies, play a crucial role in regulating stellar mass growth and suppressing star formation by heating or expelling gas from the galaxy. Figure \ref{fig:AGNFeedback} shows the evolution of the ratio $\log(\dot{M}) /\log(M_\star)$, where $\dot{M}$ is the sum of the instantaneous accretion rates of all black holes in a subhalo, alongside the stellar mass growth of central galaxies. Additionally, Figure \ref{fig:Mergers} illustrates the number of mergers as a function of redshift, providing additional insights into the growth of stellar mass in central galaxies. In both cases, the analysis focuses on the most massive central galaxy from each stellar mass bin. For low-mass galaxies, supernova feedback is the dominant mechanism, regulating star formation efficiently and overshadowing the impact of AGN feedback. This distinction is evident in Figure \ref{fig:AGNFeedback}, which shows that the stellar mass growth of low-mass galaxies exhibits a more linear evolution over cosmic time,  reflecting the comparatively minor influence of AGN activity. In contrast, for massive central galaxies ($M_\star > 10^{10.25}$ \Msunh), the AGN feedback becomes a dominant factor in the regulation of stellar mass growth. This is evident in Figure \ref{fig:AGNFeedback}, where the peaks in $\log_{10}(\dot{M})/\log_{10}(M_\star)$ (red dashed line) correspond to phases of high black hole accretion rates. These peaks indicate periods of intense AGN activity, which reduce the rate of stellar mass growth, further emphasizing the pivotal role of AGN feedback in suppressing star formation and shaping the evolutionary trajectories of massive galaxies.

On the other hand, Figure \ref{fig:Mergers} examines the number of merger events as a function of redshift for central galaxies in each stellar mass bin. The figure demonstrates that, as the stellar mass increases, the number of mergers also increases, particularly at higher $z$. This finding indicates that mergers are a dominant factor in the assembly of more massive galaxies, driving their rapid growth and intensifying the activity of their central black holes.

These findings collectively provide strong evidence that AGN feedback and mergers play a significant role in galaxy evolution, particularly for massive galaxies. This influence extends beyond what can be explained by halo formation time, supporting the conclusion that galaxy evolution is governed by multiple interconnected processes.

\section{Mutual Information (MI)}
\label{sec:ApenMI}
For continuous random variables $x$ and $y$, the MI is defined as follows \citep{shannon1948mathematical,10.5555/1592967}:
\begin{equation}
MI(x,y)= \int_{x} \int_{y} P(x,y) \mathrm{ln} \left(\frac{P(x,y)}{P(x)P(y)}\right) dxdy.
\label{eq:mutualConti}
\end{equation} 
 
Here, $P(x,y)$ represents the joint probability density function, while $P(x)$ and $P(y)$ are the marginal probability density functions, and the logarithmic term ($\mathrm{ln}$) is used to measure the MI in natural units ($nats$)\footnote{In information theory, the fundamental unit of information is the bit. However, for mathematical convenience, we employ the $nats$, which is defined as the amount of information contained in an event with a probability of $1/e$. $1 nat = 1/\mathrm{ln}(2) bits$.} The MI quantifies the reduction in uncertainty of variable $y$ based on the knowledge of variable $x$. If $x$ and $y$ are independent, $MI(x,y)=0$ \citep{pnas.1309933111}. However, estimating the MI presents a significant challenge as it requires the accurate estimation of $P(x,y)$ (see \citet{2015arXiv150907577V} for a comprehensive review). 

In our analysis, $P(x,y)$ in the equation (\ref{eq:mutualConti}) represents the joint probability distribution estimated using GMM-MI, where $x$ represents a galaxy property (assembly time, colour, sSFR, or $F_\star$) within a specific stellar mass bin and $y$ represents the assembly time of the host halo. To estimate the MI and its associated uncertainty, GMM-MI \citep{Piras23} combines Gaussian Mixture Models (GMMs) with a bootstrap technique. GMMs are probabilistic models used to describe data sets generated from a combination of multiple Gaussian distributions. Their importance lies in the ability to adjust $x$ and $y$ or the joint distribution to multiple Gaussian components. This allows GMM-MI to better capture the underlying structure of the data compared to using a single Gaussian distribution, leading to a more robust estimation of the joint probability function needed to calculate MI. Compared to traditional methods that rely on binning data and summing over joint histograms, the main advantage of GMMs is the ability to fit the x and y distribution or joint probability to multiple Gaussian components capturing the underlying continuous structure of the data, even when the true distributions might not perfectly align with chosen bin sizes. To estimate the uncertainty of the MI calculation, GMM-MI uses a bootstrap technique. It works by creating multiple new datasets (called pseudo-datasets) by randomly sampling (with replacement) from the original data. By calculating the MI in each of these pseudo-datasets, GMM-MI can assess the variability of the MI estimate and quantify the uncertainty associated with the final result. Another parameter in GMM-MI is the number of Gaussian components. To determine the number of Gaussian components for each galaxy or halo property, we used in GMM-MI the validation loss based on the log-likelihood method. This method prioritizes selecting components that lead to a model performing well on unseen data, helping ensure the estimated MI is more generalizable. Alternative information criteria such as the Akaike Information Criterion (AIC) or the Bayesian Information Criterion (BIC) can also be used in GMM-MI, which penalizes models for complexity, for selecting the optimal number of components (see details in \citet{Piras23}). These might be explored in future work for potentially more balanced component selection.

\end{document}